\documentclass[reprint,aps,pre,superscriptaddress,noeprint]{revtex4-1}
\usepackage{amsmath}
\usepackage{amsthm}
\usepackage{amsfonts}
\usepackage{bm}
\usepackage{enumitem}
\usepackage{graphicx}
\graphicspath{{figure/}}
\usepackage{epstopdf}
\usepackage{longtable}
\usepackage{hyperref}

\newcommand{\pderiv}[2]{\frac{\partial {#1}}{\partial {#2}}}

\let\originalleft\left
\let\originalright\right
\renewcommand{\left}{\mathopen{}\mathclose\bgroup\originalleft}
\renewcommand{\right}{\aftergroup\egroup\originalright}

\begin{document}

\title{Image Inversion and Uncertainty Quantification for Constitutive Laws of Pattern Formation}

\author{Hongbo Zhao}
\affiliation{Department of Chemical Engineering, Massachusetts Institute of Technology \\ 77 Massachusetts Avenue, Cambridge, MA 02139}
\author{Richard D. Braatz}
\affiliation{Department of Chemical Engineering, Massachusetts Institute of Technology \\ 77 Massachusetts Avenue, Cambridge, MA 02139}
\author{Martin Z. Bazant}
\affiliation{Department of Chemical Engineering, Massachusetts Institute of Technology \\ 77 Massachusetts Avenue, Cambridge, MA 02139}
\affiliation{Department of Mathematics, Massachusetts Institute of Technology \\ 77 Massachusetts Avenue, Cambridge, MA 02139}
\date{\today}

\begin{abstract}
The forward problems of pattern formation have been greatly empowered by extensive theoretical studies and simulations, however, the inverse problem is less well understood. It remains unclear how accurately one can use images of pattern formation to learn the functional forms of the nonlinear and nonlocal constitutive relations in the governing equation. We use PDE-constrained optimization to infer the governing dynamics and constitutive relations and use Bayesian inference and linearization to quantify their uncertainties in different systems, operating conditions, and imaging conditions. We discuss the conditions to reduce the uncertainty of the inferred functions and the correlation between them, such as state-dependent free energy and reaction kinetics (or diffusivity). We present the inversion algorithm and illustrate its robustness and uncertainties under limited spatiotemporal resolution, unknown boundary conditions, blurry initial conditions, and other non-ideal situations. Under certain situations, prior physical knowledge can be included to constrain the result. Phase-field, reaction-diffusion, and phase-field-crystal models are used as model systems. The approach developed here can find applications in inferring unknown physical properties of complex pattern-forming systems and in guiding their experimental design.
\end{abstract}

\maketitle

\section{Introduction}
Beyond the aesthetic value of pattern formation widely observed in many systems \cite{Cross1993}, their images can be harnessed to distill useful physical properties and test theoretical models quantitatively. In contrast to natural images and neural network models, images of pattern formation often lie on a manifold that can be described by relatively simple partial differential equations (PDE), despite the high degrees of freedom at the microscopic scale \cite{Turing1990} and the richness of the pattern itself \cite{Pearson1993}. Identifying the PDEs that govern an evolving pattern helps to uncover its mechanism and the constitutive relations with a small number of experiments \cite{Zhao2020}. Physically, the solution to the inverse problems reveals nonequilibrium behaviors that are difficult to compute from first principles, providing a data-driven approach to modeling complex behavior. However, due to the nonlinearity of the PDEs, it remains unclear to what extent the constitutive laws of pattern formation, especially the uncertainty in the functional form of the nonlinear and nonlocal dependence, can be reliably and robustly identified from images.

Analytical approaches have yielded valuable insights into the scaling laws of pattern formation \cite{Deen2011,Fraggedakis2020} and hence how the dynamics depend on key parameters. For spinodal decompositions, it is well known that the characteristic wavenumber of the initial spinodal pattern is related to the second derivative of the free energy \cite{CahnJohnE.1958}, and when the coarsening is governed by diffusion, the domain size grows as $t^{1/3}$ in time \cite{Furukawa1985,Bray2002,Lifshitz1961}. The growth law is found to vary with different models of concentration-dependent diffusivity\cite{Manzanarez2017}. For nonconserved fields, Ginzburg-Landau theory predicts a well-known $t^{1/2}$ growth law \cite{Lifshitz1961,Bray2002}. The structure factor of the pattern at different stages of the coarsening in both simulations and experiments has been found to collapse with the appropriate scaling \cite{Furukawa1985,Nishi1975}.

However, scaling analysis does not describe the morphology in real space. The growth law only applies when the domain is sufficiently large and no boundary effects are present \cite{Bray2002,Lifshitz1961}. In addition, for systems with competing dynamics such as reaction and diffusion, the pattern depends sensitively on both reaction kinetics and diffusivity \cite{Kondo2010,Pearson1993}, and does not necessarily follow a simple growth law. Hence, in this work, we identify the governing equation from images through PDE-constrained optimization and use uncertainty quantification to assess the accuracy of the constitutive relations.

Previous studies have demonstrated the possibility of inferring constitutive relations or governing equations from images \cite{Zhao2020,Bukshtynov2011,Bukshtynov2013,Sethurajan2015,Sethurajan2019,MoralesEscalante2020,Borisevich2012,Li2020,You2020}. Parameter identification (including the nonlocal interaction kernel) and model selection in phase-separating systems has also been studied in detail \cite{Hawkins-Daarud2013,Kahle2019,Rocca2020,Zhao2020}. With images, we have the opportunity to refine simplified models and allow unknown closures in the governing equation to be informed by data. However, for pattern-forming systems, the uncertainty in the functional form of the nonlinear or nonlocal constitutive relations remains largely unknown. Similarly, the sensitivity of the pattern morphology to the functional form is unclear, and cannot be easily elucidated by the scaling analysis mentioned above. In addition, while the patterns can be strongly influenced by specific conditions such as the average volume fraction and reaction rate, a systematic analysis of inversion based on images obtained under different conditions is lacking. We discuss solutions to these issues above in this work.

Recently, it is found that the low-dimensional manifold learned from images of spinodal decomposition is well correlated with the free energy barrier and average concentration \cite{Schoeneman2019,Schoeneman2020}. Here, we take the inverse problem approach; instead of creating a forward mapping from physical properties to patterns, we ask what physical properties can be accurately inferred from full images. In contrast with learning from the low-dimensional representation of the images (which more broadly can be obtained from feature engineering \cite{Severson2019} and dimensionality reduction techniques \cite{Hinton2006}), we utilize the full dataset in defining the objective function and in constructing the likelihood model to maximize the retrieval of useful information.

As with any regression problem, care must be taken with regards to generalization and extrapolation. Regularization is needed for the inversion of functions (infinite-dimensional inverse problems) to ensure the problem is well-posed. For example, neural networks have been used to discover or represent physical laws and constitutive relations by incorporating physical constraints \cite{Raissi2020,Raissi2019,Raissi2018,Tipireddy2019,Tartakovsky2018,Greydanus2019,Cranmer2020,Huang2020,Xu2021,Linka2021}. Sparse or symbolic regression are also commonly used to achieve parsimony and better physical interpretability \cite{Rudy2017,Long2019,Brunton2016,Maddu2019,Udrescu2020,Voss2004}. Physical constraints such as force equilibrium have been integrated with material properties to enable data-driven solutions to problems in mechanics either by using data directly without a model or inverting the constitutive model from data\cite{Kirchdoerfer2016,Ibanez2017}. Our approach enforces the general form of the governing equation while allowing the unknown dynamics (such as reaction and diffusion) to be identified. We also allow the unknown constitutive relations to be nonlinearly or nonlocally dependent on the order parameters. Additionally, imposing symmetry in the constitutive relations and other prior knowledge such as the miscibility gap can narrow down the uncertainty and prevent overfitting.

Recently, surrogate models such as Gaussian processes \cite{Wu2018}, polynomial chaos expansion \cite{Owen2017,Yan2019}, deep neural networks \cite{Tripathy2018,Yan2020}, generative adversarial networks \cite{Yang2019}, and physics-informed neural networks \cite{Zhang2019} have been used in inverse uncertainty quantification and they are particularly useful when the model evaluation is expensive. Here, we choose the full Bayesian approach \cite{Tarantola2005,Stuart2010,Sethurajan2019} based on the full PDE model and given snapshots to infer the posterior distribution of the constitutive relations. To address the curse of dimensionality that is associated with the large number of parameters used to represent functions, we use a dimension-independent Markov Chain Monte Carlo \cite{Cui2016} to ensure efficient sampling.

Our method offers a top-down approach to the construction of constitutive relations of complex systems directly from macroscopically observed fields. It is complementary to the bottom-up multiscale simulation approaches of scale bridging and learning closure models \cite{Kevrekidis2003,Kevrekidis2004,E2003,Hana2019}. With advanced imaging capabilities and the availability of more image data, the top-down data-driven approach to modeling pattern formation becomes increasingly relevant and hence calls for a detailed analysis of how well constitutive laws can be extracted from data \cite{Kalinin2015,Kalinin2016}.

We formulate the methods for solving PDE-constrained optimization and uncertainty quantification in Section \ref{sec::method} and apply the methods to various examples in Section \ref{sec::application} to study the inversion result, uncertainty and correlation in the inferred quantities, and whether models can be distinguished. In Section \ref{sec::opt_and_UQ}, the convergence of the optimization is demonstrated; the uncertainty quantification is applied to phase field and phase field crystal models to illustrate the quality of the sampling. We study phase-separating systems driven by diffusion and/or reaction as they relax toward equilibrium under different average concentration (Section \ref{sec::starting_composition}), identify the contribution of reaction and/or diffusion (Section \ref{sec::model_selection}), and when system is chemically driven out of equilibrium (Section \ref{sec::autocatalysis}). We also study the effect of imaging conditions including temporal resolution (Section \ref{sec::temporal_res}), spatial resolution (Section \ref{sec::space_res}), image domain size (Section \ref{sec::domain_size}), as well as blurring (Section \ref{sec::blurring}) on the inversion results and corresponding uncertainties.

\section{Method} \label{sec::method}
We study a class of pattern-forming systems that has a non-convex free energy $F[c]$ as a functional of the order parameter field $c(x)$.
Phase separation occurs due to instability, that is, there exists $\delta c$ such that $\delta^2 F<0$. When the order parameter is a conserved parameter such as concentration, the dynamics can be driven toward equilibrium by diffusion,
\begin{equation} \label{eqn::Cahn-Hilliard}
  \pderiv{c}{t} = \nabla \cdot \left( D(c) c \nabla \mu \right),
\end{equation}
where the functional derivative $\mu = \delta F / \delta c$ is the chemical potential.
In chemically driven systems where the interior of the system is in direct contact with the chemical reservoir, as found in surface adsorption and surface-reaction-limited nanoparticles, and when the driving is small enough to use linear irreversible thermodynamics \cite{Bazant2013,Kondepudi2014}, the governing equation is
\begin{equation} \label{eqn::Allen-Cahn}
  \pderiv{c}{t} = - R_0(c) \mu.
\end{equation}
Being in contact with a chemical reservoir, the free energy is now $F[c] - \mu_\text{res}\int{c dx}$ with the addition of a Lagrange multiplier. $\mu_\text{res}$ is the chemical potential of the chemical reservoir.

Thermodynamically, the region of instability $\delta^2 F<0$ is prohibited unless the system is out of equilibrium. Therefore, the temporal evolution of the patterns gives us access to constitutive relations in the unstable region. In the case of a phase-separating system, the free energy is typically described by a regular-solution (or Ginzburg-Landau) type double-well energy with a gradient penalty term,
\begin{equation} \label{eqn::free_energy}
  F[c(\mathbf{r})] = \int{\left( g_h(c(\mathbf{r})) + \kappa |\nabla c|^2 \right) d\mathbf{r}},
\end{equation}
where $\mathbf{r}$ is the position in space, $g_h(c)$ is the volumetric free energy when the concentration field $c(\mathbf{r})$ is homogeneous, $\kappa$ is the coefficient for penalizing concentration gradient, which leads to a diffuse interface when the system phase separates\cite{CahnJohnE.1958}. When coupled with diffusion as in Eq.\ \ref{eqn::Cahn-Hilliard}, this expression becomes the well-known Cahn-Hilliard equation used extensively in phase field models \cite{Chen2002}. When coupled with Eq.\  \ref{eqn::Allen-Cahn}, it is known as the Allen-Cahn equation.
In this case, we attempt to learn the double-well free energy $g_h(c)$ as well as the dynamics (diffusivity, and reaction kinetics). We define the homogeneous chemical potential to be $\mu_h(c) = d g_h(c) / dc$. Hence the chemical potential is
\begin{equation}
  \mu = \mu_h(c) - \kappa \nabla^2 c.
\end{equation}

Another type of pattern-forming system has a more generic nonlocal form,
\begin{equation} \label{eqn::PFC}
  F[c(\mathbf{r})] = \int{ g_h(c(\mathbf{r}))  d\mathbf{r}} + \int{ c(\mathbf{r}') C_2(|\mathbf{r}'-\mathbf{r}|)c(\mathbf{r}) d\mathbf{r} d\mathbf{r}'}.
\end{equation}
The conserved diffusion equation Eq. \ref{eqn::Cahn-Hilliard} combined with Eq. \ref{eqn::PFC} is also known as the phase field crystal equation (or dynamic density functional theory)\cite{Emmerich2012}. A particular case of a non-conserved Eq. \ref{eqn::Allen-Cahn} combined with Eq. \ref{eqn::PFC} is known as the Swift-Hohenberg equation \cite{Emmerich2012,Swift1977}.
Different direct correlation functions $C_2$ can give rise to spatial patterns such as lamellar and crystal lattice structures. In this case, we are interested in the sensitivity of the pattern with respect to the direct correlation function.

Multiple snapshots taken in time are used as training data, while the first image is used as the initial condition. Suppose the measurement noise of the images is an additive Gaussian white noise, $c_\text{data}(t_j,\mathbf{r}')-c(t_i,\mathbf{r};\mathbf{p}) \sim \mathcal{N}(0,\sigma^2 \delta(t_i-t_j,\mathbf{r}-\mathbf{r}'))$, where $c$ is the model prediction and $c_\text{data}$ is the observed concentration field, $t_i$ is time, and $\mathbf{p}$ are the parameters for the unknown constitutive relations. The conditional probability of the observed data, aka the likelihood, is
\begin{equation}
  P(c_\text{data}|\mathbf{p}) = \exp{-\frac{1}{2\sigma^2} \left[\sum_{i=1}^{M}{\int{d\mathbf{r} \, \left( c(t_i,\mathbf{r}; \mathbf{p}) - c_{\text{data}}(t_i,\mathbf{r}) \right)^2}} \right]}.
\end{equation}
where $M$ is the total number of images in time.
Similarly, we can also define the likelihood when the observed data is discrete in space and the noise is spatially and temporally uncorrelated, where the integral becomes a summation. From a Bayesian perspective, the posterior distribution of the unknown parameters $\mathbf{p}$ satisfies
\begin{equation} \label{eqn::posterior}
  P(\mathbf{p}|c_\text{data}) \propto P(c_\text{data}|\mathbf{p}) P(\mathbf{p}),
\end{equation}
where $P(\mathbf{p})$ is the prior probability, which depends on the prior knowledge of the constitutive relations discussed in detail below.

For binary mixtures, we express the free energy as a sum of the ideal entropy of mixing (id) to limit the concentration within $[0,1]$ and a non-ideal excess part (ex). The corresponding chemical potential is
\begin{equation} \label{eqn::mu}
  \mu_h(c) = \mu_\text{id}(c) + \mu_\text{ex}(c) = \ln{\frac{c}{1-c}} + \sum_{n=1}^N{a_n P_n(c)},
\end{equation}
where the $P_n$ are normalized Legendre polynomials defined on the interval [0,1], and $a_n$ are the coefficients to be determined. Similarly, the diffusivity (or reaction kinetic prefactor) can be parameterized as $\ln{D(c)} = \sum_n{b_n P_n(c)}$ to ensure positivity.
Assuming the prior for the non-ideal part of the (excess) chemical potential follows a Gaussian distribution $\mu_\text{ex}(c) \sim \mathcal{N}(0,\delta(c-c'))$, or similarly if the prior for diffusivity is $\ln{D(c)} \sim  \mathcal{N}(0,\delta(c-c'))$, then due to the orthonormality of $P_n$,
\begin{equation} \label{eqn::prior_L2}
  P(\mathbf{p}) \propto \exp{ - \frac{1}{2} \| \mathbf{p} \|_2^2 },
\end{equation}
where $\mathbf{p}$ are the coefficients $a_n$. The covariance for the priors may be defined by differential operators to penalize high-frequency components in the constitutive relations. For example, $\ln{D(c)} \sim  \mathcal{N}(0,-\mathcal{L}^{-1})$, where $\mathcal{L}\psi(c) = \frac{d}{dc} \left[ c(1-c) \frac{d}{dc} \psi(c) \right]$. In other words, $p(\psi(c)) \propto \exp{\left[ - 0.5\int{\psi(c)L\psi(c)dc}\right]}$.
Legendre polynomials (defined on [0,1]) are eigenfunctions of the differential operator, $\mathcal{L}P_n(c) = -n(n+1)P_n(c)$. Hence,
\begin{equation} \label{eqn::prior_Legendre}
  P(\mathbf{p}) \propto \exp{ -\frac{1}{2} \sum_{i=0}^\infty{n(n+1) p_i^2}}.
\end{equation}
Note that the coefficient for the constant term $P_0(c)$ is 0 (degenerate in $p_0$).

The direct correlation function in Eq.\  \ref{eqn::PFC} is represented in Fourier space in the form
\begin{equation} \label{eqn::hermite}
  \hat{C}_2(\mathbf{k}) = \sum_{n=0}^N{d_n ( 2^n n! \sqrt{\pi})^{-1/2} e^{-|\mathbf{k}|^2/2} H_n(|\mathbf{k}|)},
\end{equation}
where $\hat{C}_2(\mathbf{k}) = \int{C_2(\mathbf{r}) e^{-i\mathbf{k}\cdot\mathbf{r}}d\mathbf{r}}$, $H_n$ is the physicists' Hermite polynomial and the basis functions are orthonormal, and $d_n$ is the corresponding coefficient. Similarly, if we assume that function $r \mapsto \hat{C}_2(r\mathbf{e})$ where $\mathbf{e}$ is a unit vector follows a Gaussian distribution with delta variance,
the prior probability distribution can be written as Eq.\  \ref{eqn::prior_L2}, where $\mathbf{p}$ are the coefficients $d_n$.

For a general prior $\mathbf{p} \sim \mathcal{N}(0,\mathbf{\Gamma}_\text{Pr})$, it is convenient to transform $\mathbf{z} = \mathbf{\Gamma}_\text{Pr}^{-1/2} \mathbf{a}$ so that $\mathbf{z} \sim \mathcal{N}(0,\mathbf{I})$, which is shown below to be useful for the Markov chain Monte Carlo (MCMC) sampling and linear constraints.

In addition to a prior that promotes smoothness and penalizes high-order basis functions, priors can be modified to satisfy certain constraints. For example, in addition to the images that capture the transient behavior, the compositions of equilibrium phases ($c_1$,$c_2$) -- also known as the miscibility gap -- can be easily accessible; hence the chemical potential is subject to the thermodynamic constraint,
\begin{align} \label{eqn::mu_constraint}
  \int_{c_1}^{c_2}{\mu_h(c) dc} &= \mu_h(c_1) (c_2 - c_1) \\
  \mu_h(c_1) &= \mu_h(c_2).
\end{align}
For the Cahn-Hilliard equation, an arbitrary constant term can be added to $\mu_h(c)$, so we can set $\mu_h(c_1)=\mu_h(c_2)=0$. These three equations are a set of linear constraints on the coefficients $\mathbf{Bz} = \mathbf{d}$. For a general prior on the coefficients, it is convenient to decompose $\mathbf{z}$ into two orthonormal spaces, $\mathbf{A}$ and $\mathbf{A}^\perp$, where $\mathbf{A}$ is the null space of $\mathbf{B}$, $\mathbf{BA} = \mathbf{0}$. Then we have $\mathbf{z} = \mathbf{A}\mathbf{\xi} + \mathbf{A}^\perp \mathbf{\xi}^\perp$. Thus the constraint is $\mathbf{BA}^\perp \mathbf{\xi}^\perp = \mathbf{d}$. It can be shown that the conditional prior is
\begin{equation} \label{eqn::prior_L2_constraint}
  p(\mathbf{z}|\mathbf{Bz}=\mathbf{d}) \propto \exp{ - \frac{1}{2} \| \mathbf{\xi} \|_2^2 }.
\end{equation}

According to Bayes theorem, the posterior of the unknown parameters is $P(\mathbf{p}|c_\text{data}) \propto P(c_\text{data}|\mathbf{p}) P(\mathbf{p})$. The maximum \textit{a posteriori} estimate (MAP) is defined by minimizing the objective function
\begin{equation}
  S(\mathbf{p}) = \frac{1}{2\sigma^2} \left[\sum_{i=1}^{M}{\int{d\mathbf{r} \, \left( c(t_i,\mathbf{r}; \mathbf{p}) - c_{\text{data}}(t_i,\mathbf{r}) \right)^2}} \right] + \frac{1}{2} \mathbf{p}^* \Gamma_\text{Pr}^{-1} \mathbf{p}
\end{equation}
Without loss of generality, 10 parameters are used for all functions studied here, that is, Legendre polynomials of order 1 to 10 are used for the chemical potential, and of order 0 to 9 are used for the diffusivity and reaction kinetics. When the thermodynamic constraint (Eq.\ \ref{eqn::mu_constraint}) is applied, Legendre polynomials of order 0 to 12 are used.

The objective function is optimized using a gradient-based optimizer. The gradient of the objective function is
\begin{equation}
  \pderiv{S}{p} = \frac{1}{\sigma^2}\sum_{i=1}^{M}{\int{d\mathbf{r} \, \left( c(t_i,\mathbf{r}; \mathbf{p}) - c_{\text{data}}(t_i,\mathbf{r}) \right) \pderiv{c}{\mathbf{p}} }} +  \mathbf{\Gamma}_\text{Pr}^{-1} \mathbf{p}.
\end{equation}
The model sensitivity $\partial c / \partial \mathbf{p}$ for each parameter can be computed while solving the forward problem. For a general PDE $\pderiv{c}{t} = g(t,c;\mathbf{p})$
\begin{equation} \label{eqn::FSA}
  \pderiv{}{t} \left(\pderiv{c}{\mathbf{p}}\right) = \pderiv{g}{c} \cdot \pderiv{c}{\mathbf{p}} + \pderiv{g}{p},
\end{equation}
which is known as forward sensitivity analysis (FSA). From the FSA, we estimate the Hessian of the objective function using Gauss-Newton approximation \cite{Fletcher2000},
\begin{equation} \label{eqn::Gauss-Newton}
  \mathbf{H}[S] \approx  \frac{1}{\sigma^2} \sum_{i=1}^{M}{\int{d\mathbf{r} \, \left( \pderiv{c}{\mathbf{p}} \right)^* \pderiv{c}{\mathbf{p}} }} + \mathbf{\Gamma}_\text{Pr}^{-1}.
\end{equation}
The approximation becomes increasingly accurate as the difference between the model and data is decreased.
An alternative to FSA is adjoint sensitivity analysis (ASA), which involves solving the adjoint linear sensitivity equation backward in time once to obtain the sensitivity of the objective function with respect to all parameters \cite{Maly1996,Cao2002,Cao2003} (equivalent to backpropagation for neural networks). When ASA is used, only the gradient is computed and the optimizer can use a gradient descent algorithm. This approach is useful for a large number of parameters, such as a 2D field or weights in neural nets. When the number of parameters is small, such as in the case of parametrizing concentration-dependent functions, the benefit of using FSA to obtain an estimate of the Hessian outweighs its computational cost, as the convergence is much faster. We use the trust-region algorithm\cite{Fletcher2000} for the optimization when the Hessian is available.

Given the posterior distribution $P(\mathbf{p}|c_\text{data})$, we quantify the uncertainty by sampling the parameter space using Markov Chain Monte Carlo (MCMC). A popular adaptive Metropolis algorithm updates the covariance of the proposal distribution adaptively from the chain samples \cite{Haario2001}, that is, the covariance at step $n$ is $C_n = s_d \text{Cov}(X_0,\ldots X_{n-1})+s_d \epsilon I$, where $\text{Cov}$ is the sample covariance from the previous $n$ samples and $s_d = 2.4^2/d$, where d is the dimensionality of the parameter space. This method does not compute the gradient or Hessian. In our case, we need an MCMC sampler for functions, which are parameterized in a finite-dimensional space by truncating the polynomials. With an increasingly fine representation of the function (higher dimension), the adaptive MCMC method becomes slow and the mixing quality of the Markov chain deteriorates rapidly with finer representations of the functions\cite{Stuart2010}. The sensitivity with respect to high-order basis functions leads to poorer sampling, and the sampler stagnates for a prolonged period of time, which is especially problematic for computationally expensive model evaluations of PDEs that we study here. Here we use a dimension-independent and likelihood-informed (DILI) MCMC \cite{Cui2016} that takes the local Hessian information and adopts an operator-weighted proposal distribution to achieve better sampling efficiency that is independent of the dimensionality of the parameter space. We start the algorithm from the optimal solution (MAP). The Hessian is computed periodically using Eq.\  \ref{eqn::Gauss-Newton} to determine the parameter subspace that are most informed by the model (likelihood-informed subspace). The posterior covariance $\mathbf{\Sigma}$ is more accurately estimated by combining the covariance of projecting the chain samples onto the likelihood-informed subspace and the prior covariance in the complement prior-informed subspace. The proposal distribution comes from the discretization of the Langevin equation,
\begin{equation}
  d \mathbf{p} = - \mathbf{\Sigma} \pderiv{S}{\mathbf{p}} d \tau + \sqrt{2\mathbf{\Sigma}} d\mathbf{W},
\end{equation}
whose stationary distribution is the posterior distribution (Eq.\ \ref{eqn::posterior}), and $\mathbf{W}$ is random noise $\langle \mathbf{W}(\tau)^* \mathbf{W}(\tau') \rangle = \delta(\tau-\tau')\mathbf{I}$. As an alternative, we adopt the proposal distribution,
\begin{equation}
  d \mathbf{p} = - \mathbf{\Sigma} \mathbf{\Gamma}_\text{Pr}^{-1} \mathbf{p} d \tau + \sqrt{2\mathbf{\Sigma}} d\mathbf{W}
\end{equation}
based on gradient descent from the prior distribution only to avoid computing the model sensitivity at every step. This expression requires $\mathbf{\Gamma}_\text{Pr}$ to be nonsingular. In practice, for a degenerate distribution such as Eq.\ \ref{eqn::prior_Legendre}, we set the variance of the degenerate component (the coefficient of the constant basis function $P_0(c)$) to be sufficiently large.
We set a maximum number of Hessian evaluations that can be performed. Therefore, the extra computational time needed for this method compared to the adaptive MCMC with the same number of samples $N_s$ becomes vanishingly small as $N_s$ increases.

\section{Applications}
\label{sec::application}
\subsection{Optimization and uncertainty quantification}
\label{sec::opt_and_UQ}
\begin{figure}
  \centering
  \includegraphics[width=0.5\textwidth]{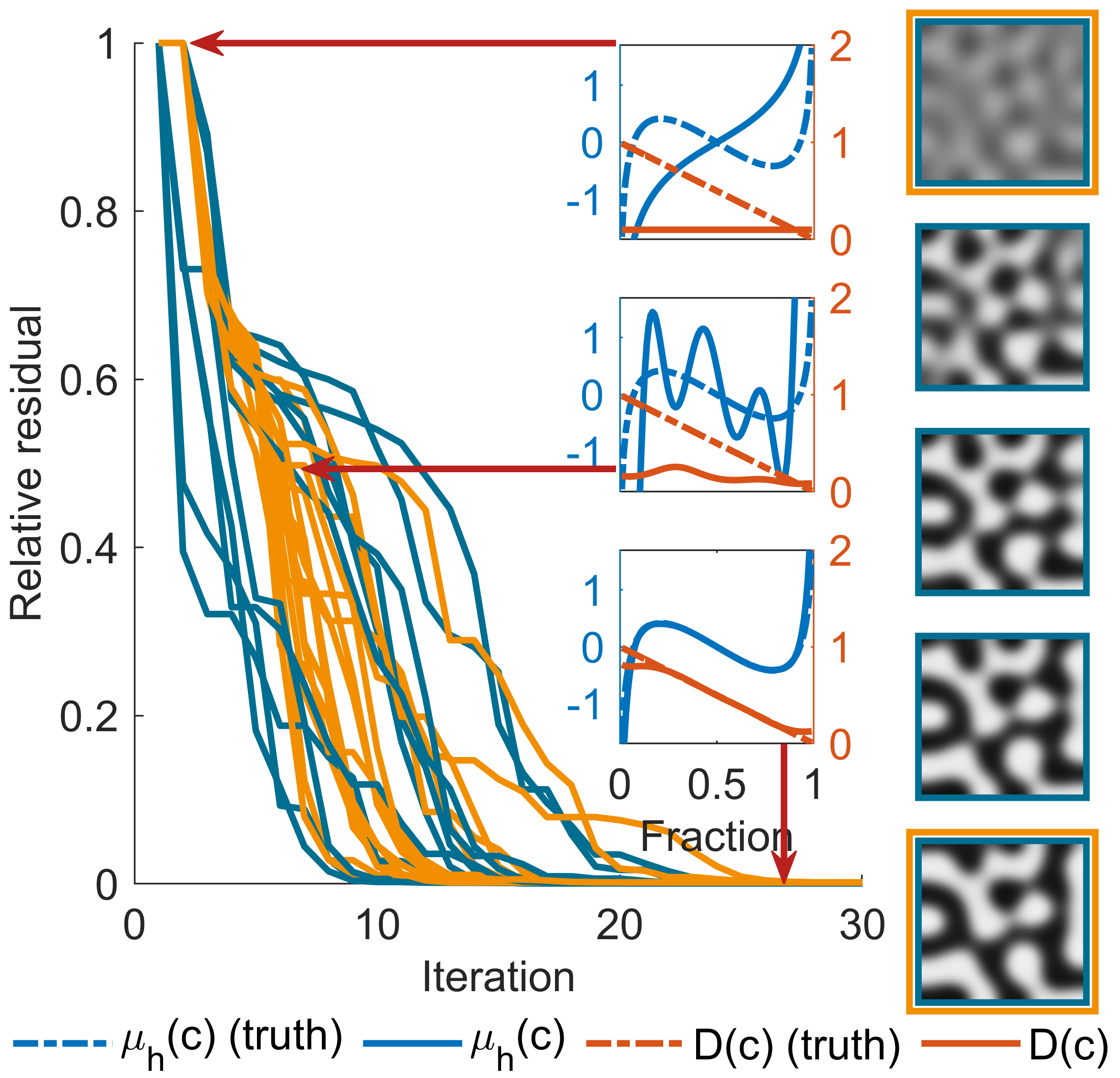}
  \caption{Training on 11 realizations of spinodal decomposition snapshots. Insets show the initial guess for $\mu_h(c)$ and $D(c)$, a typical state during the training, and its final convergence to the truth. Blue and orange curves in the main plot correspond to taking 5 and 2 snapshots as the training datasets, respectively, as highlighted by the outline of an example set of images on the right. The images are concentration fields $c(x)$ with black and white corresponding to $c=1$ and $c=0$, respectively. The same colormap is used throughout this paper.}
  \label{fig::ensemble}
\end{figure}

We generate simulated data for the evolution of the Cahn-Hilliard equation. The boundary conditions are $\mathbf{n}\cdot \nabla c=0$, which corresponds to no surface wetting, and $\mathbf{n}\cdot \nabla \mu=0$, which corresponds to no flux\cite{Bazant2013,Cahn1961}. Fig.\ \ref{fig::ensemble} shows the evolution of the objective function during the optimization MAP using different realizations of spinodal decomposition snapshots as training data, which are generated with the same image resolution and physical parameters and different random initial conditions. These results demonstrate the robustness of the optimization algorithm. Using FSA, the Gauss-Newton approximation, and the trust-region algorithm, the optimizer converges to the truth within tens of iterations robustly with 5, or as few as 2 snapshots.

10 parameters are used for both $\mu_\text{ex}$ and $\ln{D(c)}$. The initial guess for $\mu_\text{ex}$ and $\ln{D(c)}$ is chosen such that it is far from the truth and generates a non-pattern-forming evolution ($\mu_h(c) = \ln{\frac{c}{1-c}} +(1-2c)$ and $D(c)/L^2=0.1$, where $L$ is the domain size (see the inset at iteration 1 in Fig.\ \ref{fig::ensemble}; these initial guesses are used throughout the text unless otherwise noted). We recommend choosing a small $D(c)$ as the initial guess to ``freeze'' the pattern. If the pattern relaxes too fast and becomes uniform beyond the first frame, the sensitivity of the pattern with respect to all model parameters approaches zero, which is known as the vanishing gradient problem in machine learning. The initial guess for the gradient penalty is $\kappa / \kappa_\text{truth} = 0.05$. The length scale of the diffuse interface is $\sqrt{\kappa_\text{truth}}=0.045L$.

To find the MAP, $L_2$ norm is used as the regularizer for $\mu_\text{ex}(c)$ and $\mathcal{L}$ norm is used for $\ln{D(c)}$, and regularization parameters for both are $10^{-5}$ (this set of parameters is used throughout unless otherwise noted). Due to the structure of Eq.\ \ref{eqn::Cahn-Hilliard}, $\mu_h(c)$ and $D(c)$ can only be determined up to a constant scale; therefore, throughout the text, we report $\mu_h(c) \kappa_\text{truth}/\kappa$ and $D(c) \kappa / \kappa_\text{truth}$. In fact, the optimizer often converges to a different $\kappa$, while the scaled quantities above are accurately recovered. Allowing $\kappa$ to vary in the optimization generally speeds up the convergence.

In the appendix \ref{sec::appendix_alg}, we compare the performance of optimization using FSA (trust-region algorithm) versus ASA (gradient descent) and found that FSA leads to a much faster convergence, lower computational cost, and higher success rate of convergence to the global minimum (as opposed to a local minimum).

\begin{figure}
  \centering
  \includegraphics[width=0.5\textwidth]{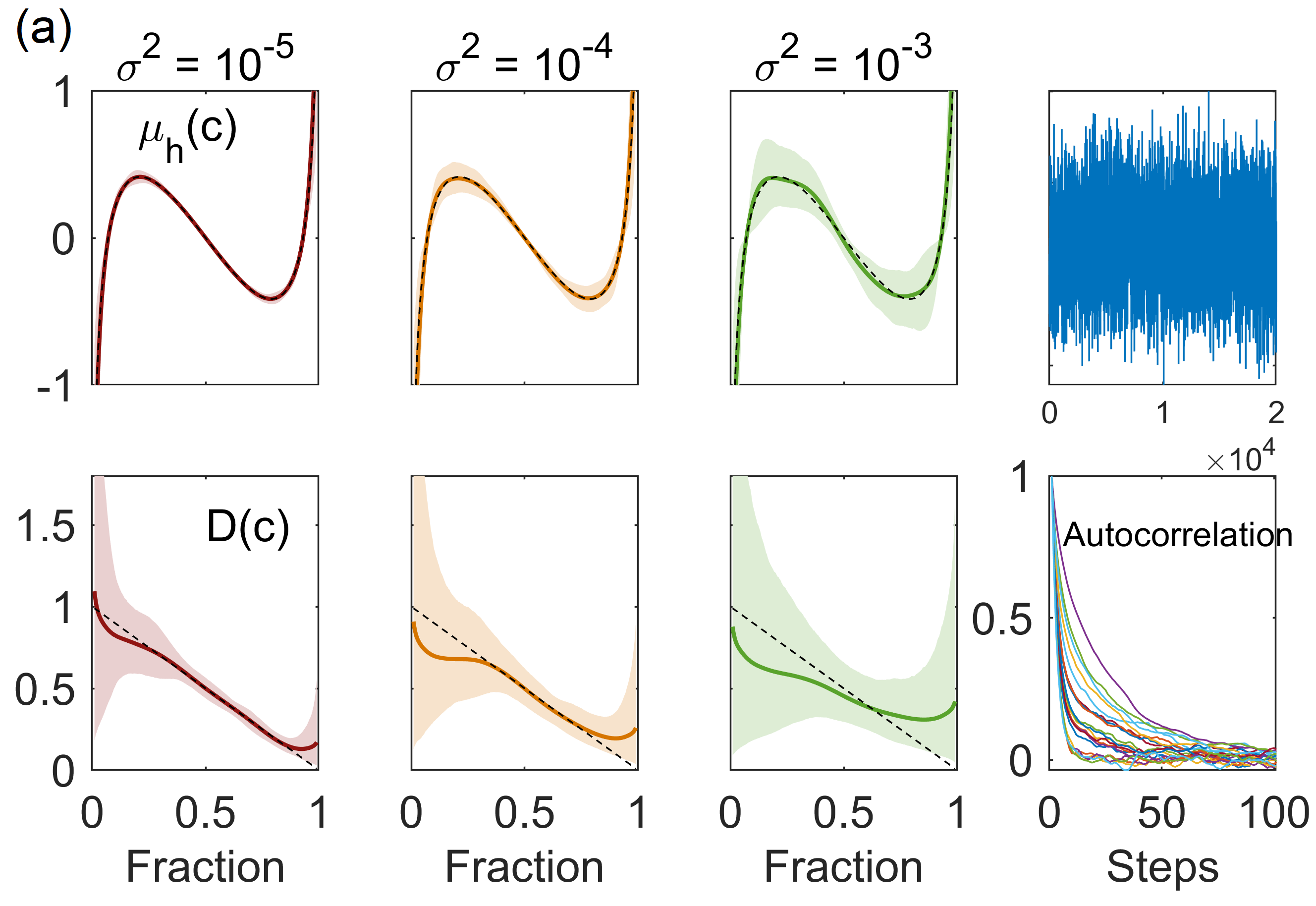}\\
  \vspace{0.5cm}
  \centering
  \includegraphics[width=0.5\textwidth]{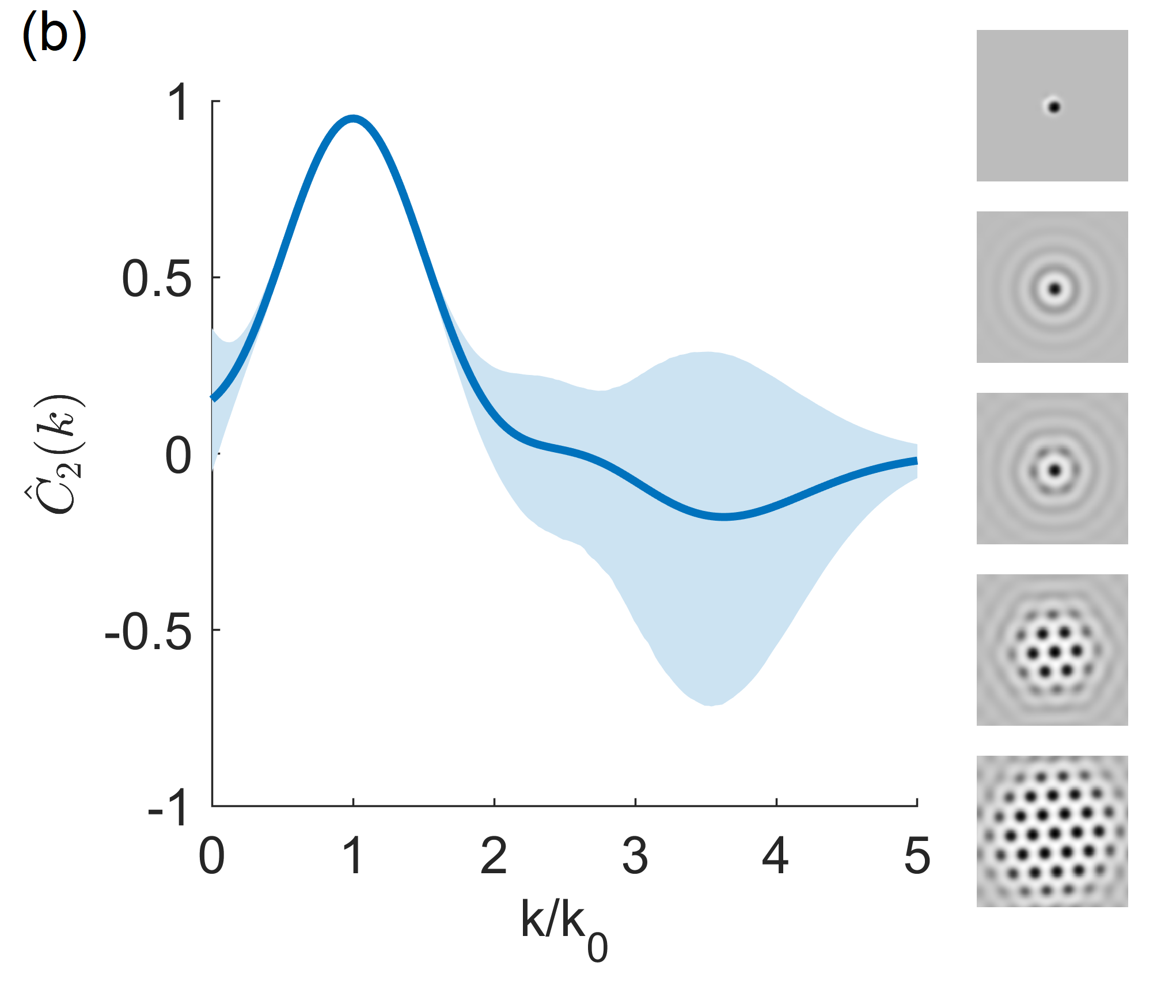}
  \caption{(a) Uncertainty quantification of the chemical potential and diffusivity from 5 snapshots of spinodal decomposition as shown in Fig.\  \ref{fig::ensemble}. The measurement noise $\sigma^2$ is varied. The shaded regions are the 95\% confidence interval of the functions at each $c$. The solid lines are the marginal mean of the functions at each $c$. The dashed lines are truth. This plotting convention is used throughout the paper unless otherwise noted. The two panels on the right are the Markov chain trajectory of $a_1$ and the autocorrelation of all 20 parameters, respectively. (b) Uncertainty quantification of the direct correlation from phase-field-crystal model. Images on the right are the training data.}
  \label{fig::MCMC}
\end{figure}

The optimization allows us to quickly find the MAP, which serves as the starting point for MCMC. For the uncertainty analysis, we assume the measurement is a continuous field. $\kappa$ is fixed (since $\mu_h(c)$ and $D(c)$ can always be rescaled, as explained above), and $\mu_h(c)$ is subject to the constraint Eq.\ \ref{eqn::mu_constraint}. The prior for $\mu_h(c)$ and $\ln{D(c)}$ are described by Eqs.\  \ref{eqn::prior_L2_constraint} and  \ref{eqn::prior_Legendre}, respectively, here and for the rest of the paper. The Markov chain is converted to $\mu_h(c)$ and $D(c)$ and the uncertainty at each $c$ is plotted at 95\% confidence level throughout the paper. Fig.\ \ref{fig::MCMC}a shows the uncertainty quantification result and the performance of the DILI MCMC algorithm. The shaded area corresponds to the confidence interval, while the solid line is the marginal mean at each $c$. The measurement noise $\sigma$ may be obtained from the knowledge of the instrument or inferred together with other parameters. From here on, unless otherwise noted, we assume $\sigma^2= 10^{-4}$ to illustrate the sensitivity more clearly.  The Markov chain achieves good mixing within less than 100 steps for all parameters (20 in total) as shown by the autocorrelation of the chain in Fig.\ \ref{fig::MCMC}a. Fig.\ \ref{fig::MCMC} also shows the trajectory of one of the parameters, which is indicative of good sampling. $2\times 10^4$ samples are used unless otherwise noted.

In comparison, the adaptive MCMC (Fig.\ \ref{fig::AM}) has a much longer autocorrelation and hence poorer mixing. The burn-in time is also much longer. Therefore, we use DILI MCMC to perform uncertainty quantification in the following examples.

If the system is non-phase-separating, $\kappa=0$, Eq.\ \ref{eqn::Cahn-Hilliard} becomes
\begin{equation}
  \pderiv{c}{t} = \nabla \cdot \left( D(c) c \cdot \mu'_h(c) \nabla c \right).
\end{equation}
Therefore, only the chemical diffusivity $D_\text{chem} = cD\mu'_h(c)$ can be inferred and no thermodynamic information can be obtained from only the concentration field. However, the extraction of $\mu_h(c)$ and $D(c)$ is possible in phase-separating systems, and are largely uncorrelated, as shown by the correlation coefficients between the parameters $a_1,a_2,\ldots,a_{10}$ and $b_0,b_1,\ldots,b_9$ in Fig.\ \ref{fig::corrcoef}. The correlation among odd/even polynomials of low order (of the same function) is strong, while the correlation between $\mu_h(c)$ and $D(c)$ is weak. The correlation between the thermodynamic and transport properties can also be quantified through certain scalars of interest. At $\sigma^2=10^{-5}$, $-\mu'_h(c_0)$ and $D(c_0)$ are weakly and negatively correlated (Fig.\  \ref{fig::mu_D_corr}). As explained in further in detail in Section \ref{sec::model_selection}, $D(c_0)\left[ \mu_h'(c_0) \right]^2$ determines the initial rate of spinodal decomposition, but information from the entire field decouples $\mu'_h(c)$ and $D(c)$.

Another quantity of interest is the interfacial tension, which is defined as the energy of a flat interface at equilibrium $\gamma=\int_{-\infty}^\infty{\left[ g_h(c)+\frac{1}{2}\kappa \left(\pderiv{c}{x}\right)^2\right]}$. Using the equilibrium condition $\mu(x)=\mu_0$ where $\mu_0$ is the chemical potential of the two equilibrium phases, we have
\begin{equation}
  \gamma = \int_{c_1}^{c_2}{\sqrt{2\kappa \Delta g_h(c)} dc}
\end{equation}
where $\Delta g_h(c) = g_h(c)-g_h(c_1)-(c-c_1)\mu_0$. At the late stage of coarsening, the domain growth rate is proportional to $\gamma D(c)$ \cite{Bray2002}. However, we find that $\gamma$ and $D(c_0)$ (or $D(c_1)$) are almost uncorrelated (Fig.\  \ref{fig::mu_D_corr}), due to information at the diffuse interface and early-stage patterns.

For the phase field crystal models (Eq.\ \ref{eqn::PFC}), previous studies have found correlation functions that generate certain crystal structures and elastic constants \cite{Greenwood2010,Jaatinen2009}. In Fig.\  \ref{fig::MCMC}, we use MCMC to determine the uncertainty in the direct correlation function inferred from a set of images of nucleation that forms hexagonal crystal structure ($\sigma^2 = 10^{-4}$). Using the parameterization given by Eq.\  \ref{eqn::hermite} and the $L_2$ norm, we find the extent that the function is allowed to vary while keeping the same pattern is very small around wavenumbers that correspond to the spatial wavelength $k_0$ observed in the image. Uncertainty  increases for wavenumbers as moving away from $k_0$ due to the insensitivity of the model to these components and the lack of Fourier components at high wavenumbers in the images provided.

\subsection{Starting composition}
\label{sec::starting_composition}
Images taken at different operating conditions can reveal information with different levels of confidence. Patterns of spinodal decomposition at different average concentration range from tortuous regions near $c_0=0.5$ to dispersed droplets near $c_0=0.3$ and $c_0=0.7$. First, an optimization is performed to find the MAP, starting from the same initial guesses as Fig.\  \ref{fig::ensemble}, convergence to the truth is achieved within 20 iterations. Then MCMC is performed to study the uncertainty in the $\mu_h(c)$ and $D(c)$. Histograms of the images show a peak near the average concentration (also the initial condition) as strong as the peak at the miscibility gap near 0 and 1. With sufficient pixel values around $c_0$ and their dynamics driven by $\mu'_h(c_0)$ and $D(c_0)$ at the average concentration  ($\mu_h(c_1)$ is fixed to be 0), the uncertainty consistently reaches a minimum at $c_0$ for both constitutive relations and all concentrations studied. Overall, images of $c=0.5$ reduce the maximum uncertainty, while away from $c=0.5$, multiple experiments may be necessary to accurately determine the functions on a larger range of $c$.

\begin{figure}
  \centering
  \includegraphics[width=0.5\textwidth]{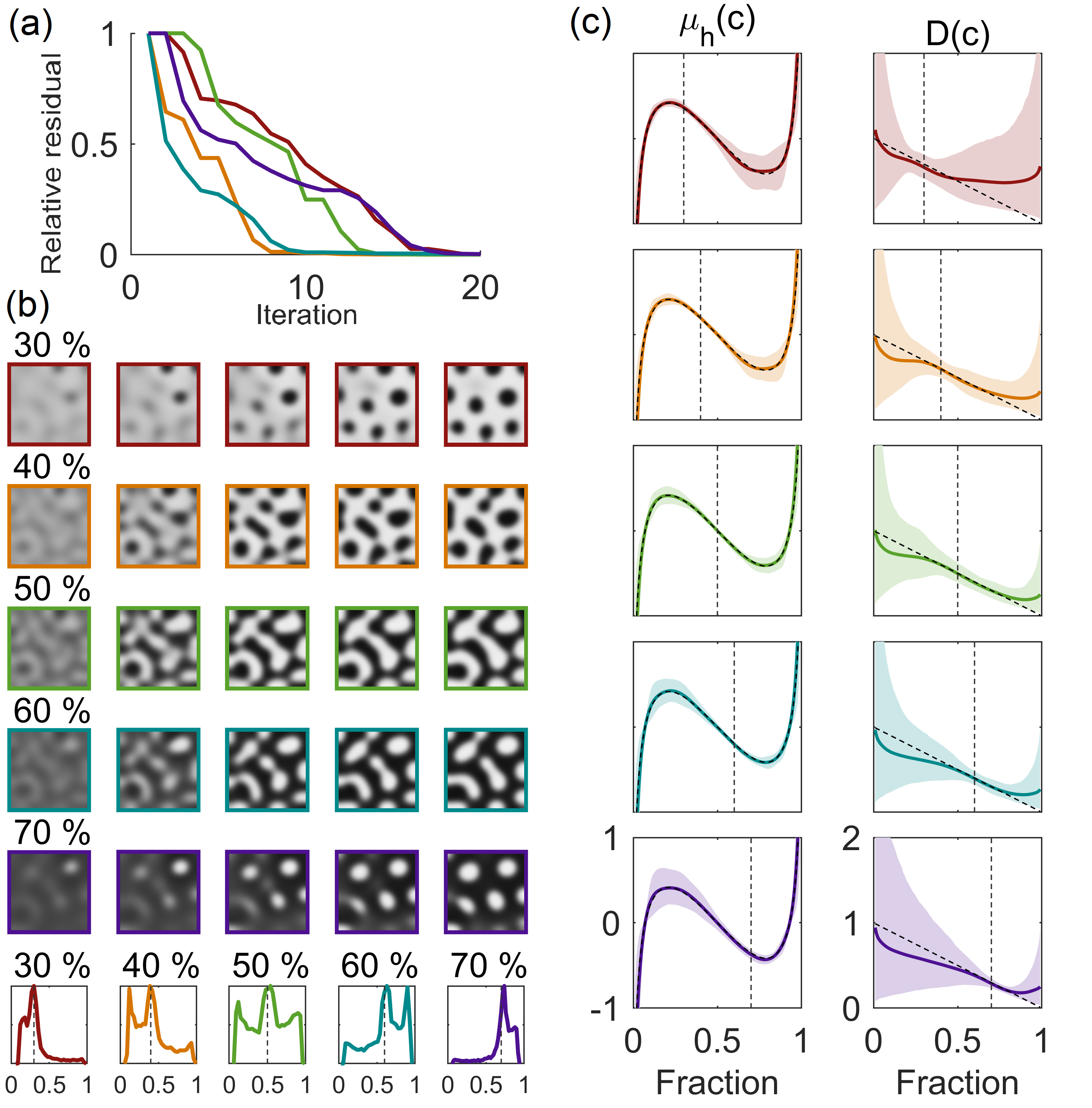}
  \caption{Training on spinodal decomposition snapshots of mixtures from 30\% to 70\%. Different composition ratios are colored consistently for all figures. (a) Relative residuals for all ratios. (b) The training datasets and their composition histograms, where the average composition is highlighted by dashed lines. (c) The uncertainty in the learned chemical potential $\mu_h(c)$ and diffusivity $D(c)$. }
  \label{fig::c0}
\end{figure}

\subsection{Model selection: Diffusion and reaction}
\label{sec::model_selection}
When the governing dynamics of the system is unknown, or it is unknown whether the order parameter is locally conserved, we consider the possibility of both conserved and nonconserved dynamics, by generalizing Eqs.\ \ref{eqn::Cahn-Hilliard} and \ref{eqn::Allen-Cahn}, and allow the magnitude of both dynamics to be inferred from the patterns,
\begin{equation} \label{eqn::CHACR}
  \pderiv{c}{t} = \nabla \cdot D(c) c \nabla \mu + R_0(c) \left(\mu_\text{res} - \mu \right).
\end{equation}
where
When solving the inverse problem, the reservoir chemical potential $\mu_\text{res}$ is unknown. We allow it to vary in time such that the average concentration $\int{cdV}/\int{dV}$ is equal to that of the images, which is constant in this case. This becomes an algebraic constraint on the model. Reaction-diffusion models have been studied extensively in literature \cite{Kondo2010,Murray2007,Wang2019}. Here we focus on a one-component system where the reaction takes place between the system and reservoir. Lithium in lithium iron phosphate (LFP) platelet particles is known to undergo diffusion in the lateral direction via a surface layer while the platelet also exchanges lithium with the electrolyte reservoir \cite{Li2014}. The dynamics of a thin platelet particle may be modeled with Eq.\ \ref{eqn::CHACR} on a 2D plane \cite{Bazant2013}.

The time scale of relaxation can be understood from the dispersion relation of Eq.\ \ref{eqn::CHACR} linearized around a homogeneous state $c(\mathbf{r})=c_0$, obtained by substituting a perturbation $e^{\omega t + i \mathbf{kr}}$ into the linearized equation  \cite{Bai2011, Bazant2013,Bazant2017}, and together with the chemical potential defined by Eq.\  \ref{eqn::free_energy},
\begin{equation}
  \omega(k) = -\left( c_0 D(c_0) k^2 + R_0(c_0) \right) \left( \mu_h'(c_0) - \kappa k^2 \right),
\end{equation}
where $k=|\mathbf{k}|$.
When diffusion dominates, $-c_0 D(c_0) \mu_h'(c_0) > R_0(c_0) \kappa$, the maximum instability growth rate is
\begin{equation}
  \omega_\text{max} = \max_{k}\omega(k) = \frac{\left( c_0 D(c_0) \mu_h'(c_0) - R_0(c_0) \kappa \right)^2}{4 c_0 D(c_0) \kappa},
\end{equation}
When there is no reaction, $\omega_\text{max} = c_0 D(c_0) \left[ \mu_h'(c_0) \right]^2 / 4\kappa$.
Otherwise, reaction dominates and
\begin{equation}
  \omega_\text{max} = - R_0(c_0) \mu_h'(c_0).
\end{equation}
When diffusion dominates, the maximum growth rate of instability is obtained at a nonzero wavenumber, while when reaction dominates, the maximum growth rate corresponds to $k=0$. Therefore, the patterns are visually different. However, when both effects are important, the relative strength of reaction and diffusion cannot be easily distinguished, which motivates a systematic approach of identifying the underlying dynamics via model selection.

\begin{figure}
  \centering
  \includegraphics[width=0.5\textwidth]{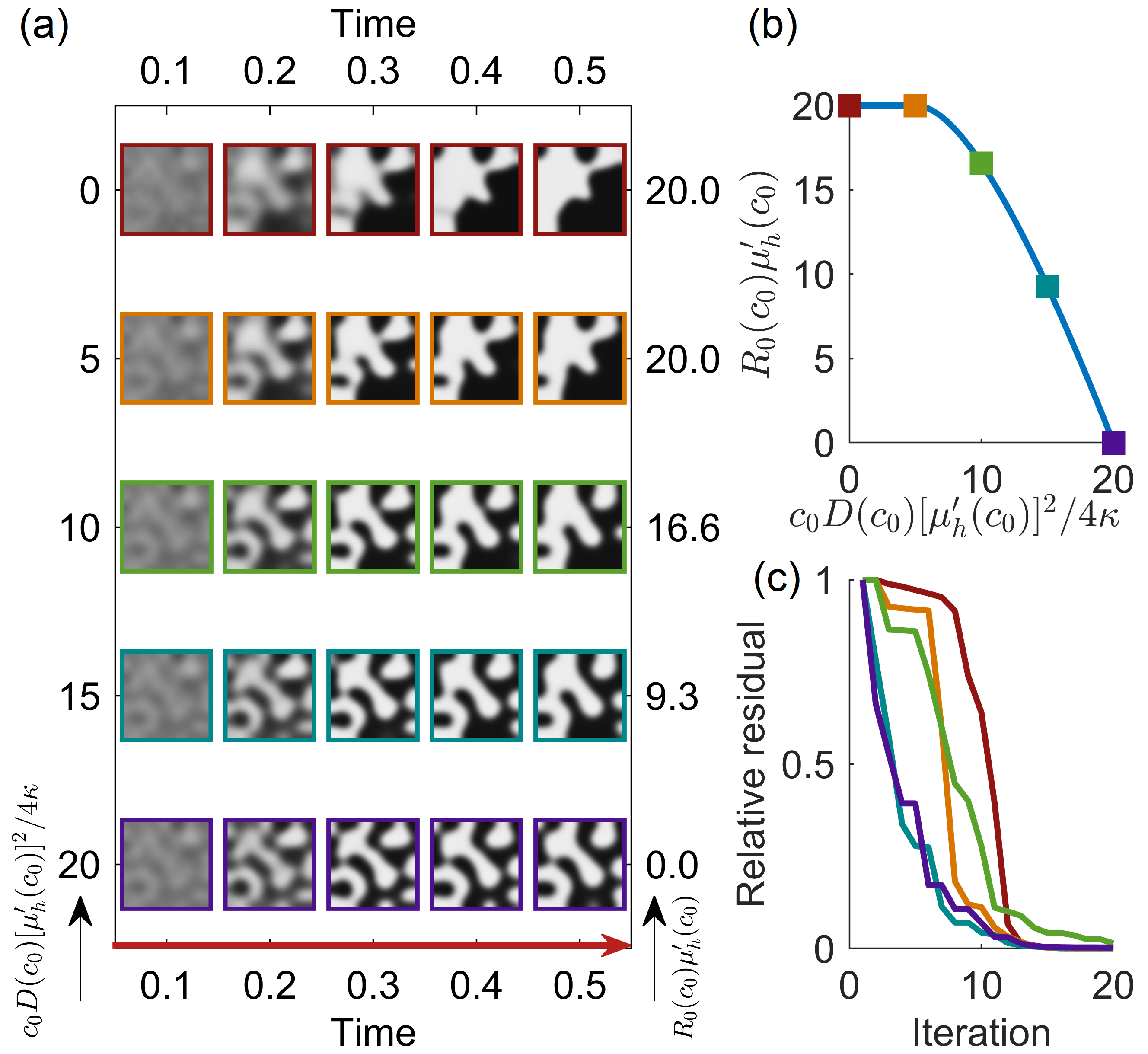} \\
  \vspace{1cm}
  \centering
  \includegraphics[width=0.5\textwidth]{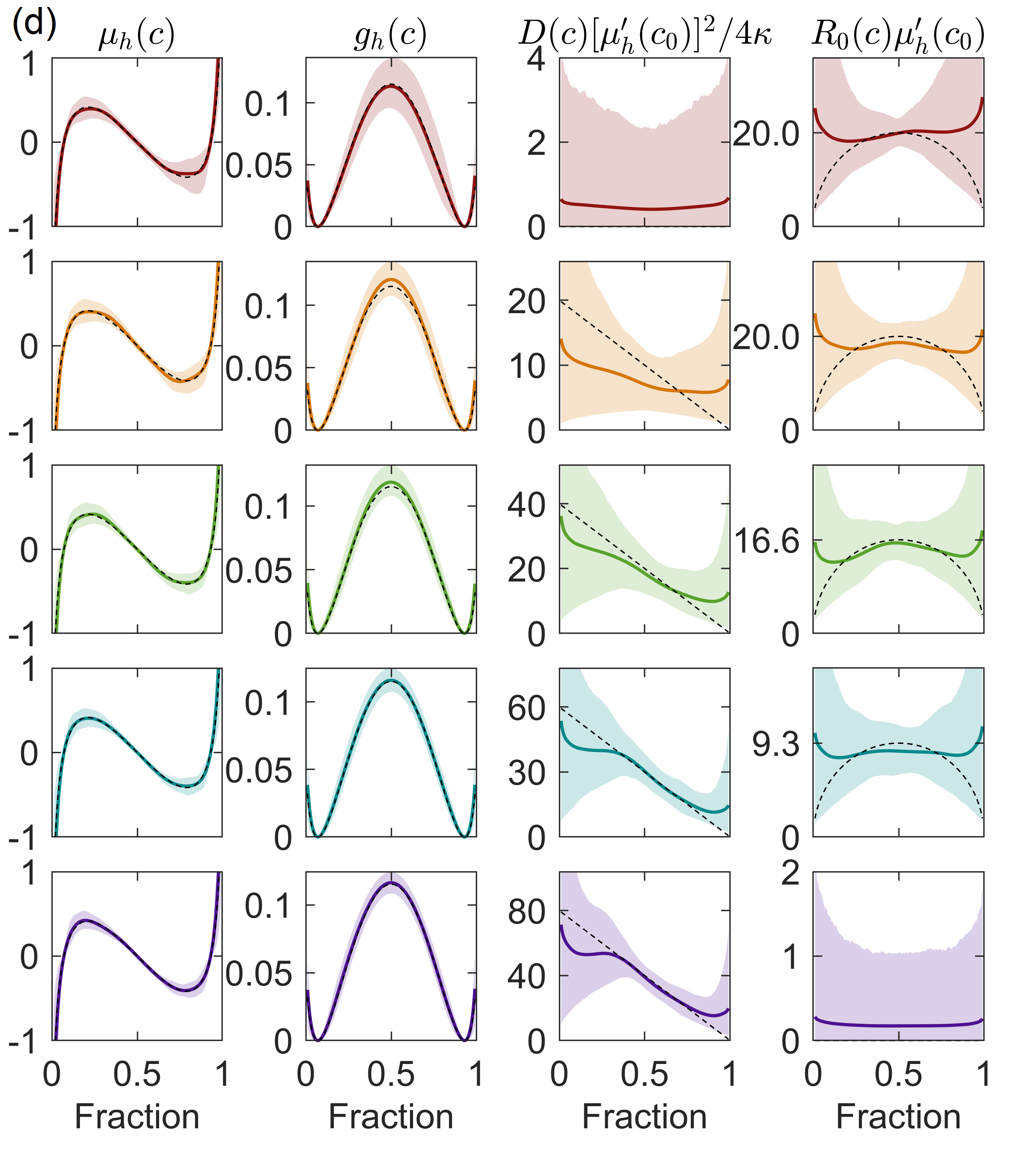}
  \caption{Dynamics of spinodal decomposition driven by varying degrees of reaction and diffusivity. The five cases studied are colored consistently in all plots. (a) The training datasets and their corresponding physical time, scaled reaction kinetic prefactor and diffusivity. (b) The relationship between scaled reaction kinetic prefactor and diffusivity with constant instability growth rate. (c) The residual plot during the training process for all cases. (d) The uncertainty in chemical potential, free energy, scaled diffusivity and reaction kinetic prefactor for all cases. The scaling constants $\mu'_h(c_0)$ and $\kappa$ are the known truth.}
  \label{fig::model_selection}
\end{figure}

We generate the images by varying the magnitude of the diffusivity while keeping the characteristic time scale $\omega_\text{max}$ constant. The iso-$\omega_\text{max}$ curve is shown in Fig.\  \ref{fig::model_selection}b. The images used as the training data for each diffusivity value, ranging from diffusion only to reaction only, are taken at the same time and shown in Fig.\ \ref{fig::model_selection}a. The chemical potential of the external reservoir adjusts passively while the total concentration is conserved. Training data and results are colored consistently based on its diffusivity in Fig.\  \ref{fig::model_selection}. The initial guess for $\mu_h(c)$ and $\kappa$ are the same as section \ref{sec::opt_and_UQ}. The initial guesses for $R_0(c)$ and $D(c)$ are $0.1$ and $0.01$ respectively for all cases. 10 parameters are used to represent each function. The residual plot Fig.\ \ref{fig::model_selection}c confirms that the truth model is found for all cases and that the images are sufficient for identifying the underlying dynamics.
When the true $D(c)$ or $R_0(c)$ is zero, the solver converges to a vanishingly small value for $D(c)$ and $R_0(c)$.

Fig.\ \ref{fig::model_selection}d shows the uncertainty in the inferred functions. The uncertainty in the chemical potential, or free energy, inferred from reaction-controlled patterns are higher than diffusion-controlled patterns, which is also reflected in the uncertainty of the interfacial tension.
For the five cases studied $D(c_0)\left[ \mu'_h(c_0) \right]^2/4\kappa = 0, 5, 10, 15, 20$, the mean and 2 standard deviation of $\gamma/\gamma_\text{truth}$ is $0.99 \pm 0.12$, $1.01 \pm 0.079 $, $1.01 \pm 0.066$, $1.00 \pm 0.057$, $1.01\pm 0.059$, respectively. The interfacial tension $\gamma_\text{truth}/\sqrt{2\kappa}$ is 0.2. At the late stage of coarsening for Allen-Cahn equation, the interface growth rate is proportional to the local curvature and independent of the interfacial tension (see Section 2.3 in Ref.\  \cite{Bray2002}), which explains the increasing uncertainty in the free energy when reaction dominates. Early-stage snapshots within the miscibility gap are essential for inferring the free energy of an Allen-Cahn system.

When reaction dominates, the diffusivity becomes increasingly uncertain, and when the true diffusivity is zero, the inferred diffusivity can be anything below a threshold. Recall that the constant term in $\ln{D(c)}$ has a degenerate prior; therefore the upper bound of $D(c)$ is determined by the likelihood only. The same is true for reaction kinetics -- when diffusion dominates, the inferred reaction kinetics becomes uncertain.  While the patterns are not sensitive to the exact form of $D(c)$ and $R_0(c)$ when either reaction or diffusion dominates, their magnitudes can be identified relatively accurately. Regardless of the magnitude of diffusivity, the uncertainty in $R_0(c)$ becomes increasingly large when $c$ approaches 0 or 1. This effect can be understood from an analysis of the sensitivity of $c(x)$ with respect to $R_0(c)$. In the sensitivity equation Eq.\ \ref{eqn::FSA}, when close to the miscibility gap, $\pderiv{}{c}
\left(\pderiv{c}{t}\right)<0$ due to thermodynamic stability, the sensitivity of the reaction rate $\delta R_0(c) \cdot \left(\mu_\text{res}-\mu(c)\right)$ and hence the sensitivity of $c(x)$ becomes increasingly small as $\mu \rightarrow \mu_\text{res}$.
Similar to Section \ref{sec::opt_and_UQ}, we observe a weakly negative correlation between $-\mu'_h(c_0)$ and $D(c_0)$ (or $R_0(c_0)$), which can be understood from their negative correlation when $\omega_\text{max}$ is constant. The correlation between $D(c_0)$ and $R_0(c_0)$ is weak (see Fig.\  \ref{fig::model_selection_corr}). The correlation among the three is at its maximum when $-c_0 D(c_0) \mu_h'(c_0) = R_0(c_0) \kappa$, which is the critical point where the system transitions from being reaction-dominated to diffusion-dominated.

\subsection{Chemically driven system: the effect of autocatalysis}
\label{sec::autocatalysis}
We study a system that is chemically driven by an external chemical reservoir at a constant average reaction rate, the concentration evolution follows \cite{Bazant2013}
\begin{equation} \label{eqn::ACR}
  \pderiv{c}{t} = R_0(c) f(\mu_\text{res}-\mu).
\end{equation}
When driven far from equilibrium, $f$ is no longer necessarily a linear function. Here we use $f(x) = 2\sinh(x/2)$, which is known as symmetric Butler-Volmer kinetics in electrochemistry \cite{Newman2012}.
The external chemical potential $\mu_\text{res}$ varies in time, subject to the constraint of the average reaction rate $R$,
\begin{equation}
  \int{\pderiv{c}{t} dV} = R\int{dV}.
\end{equation}
The linear stability and pattern formation of such a chemically driven system have been studied extensively \cite{Bazant2013,Bazant2017,Cogswell2012,Zhao2019,Fraggedakis2020a}. The dispersion relation of Eq.\ \ref{eqn::ACR} is $\omega = R_0'f - R_0 f' \cdot (\mu'_h - \kappa k^2)$. Depending on the magnitude and direction of the reaction rate $R$, the state-dependent $R_0(c)$ can alter the linear stability (and hence the pattern) to deviate from its thermodynamic stability as determined by $\mu'_h(c)$.

\begin{figure}
  \centering
  \includegraphics[width=0.5\textwidth]{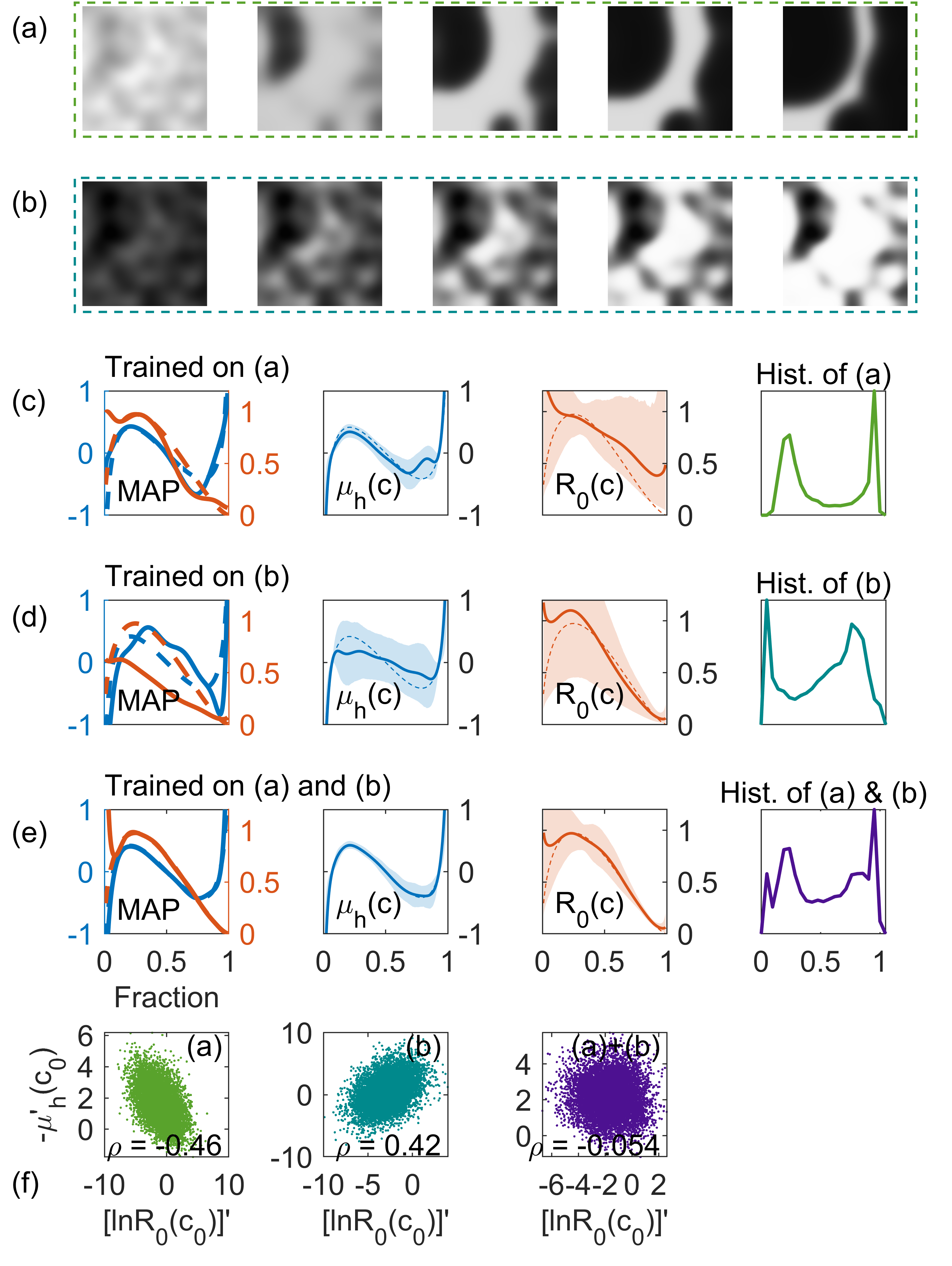}
  \caption{Learning reaction kinetics and free energy from the spatial patterns of a system chemically driven at a constant average reaction rate. (a,b) Training data where $R=0.08$ and $R=-1$, respectively. (c,d) Training result based on datasets (a) and (b), respectively. A comparison between the known truth and MAP, the uncertainty of $\mu_h(c)$ and $R_0(c)$, and the histogram of datasets (a) and (b). (e) Training result and uncertainty based on both (a) and (b) and histogram of the both datasets combined. (f) The scatter plots of $\left[\ln{R_0(c_0)}\right]'$ and $-\mu_h'(c_0)$ from the MCMC sample based on datasets (a), (b) and both combined, $c_0=0.5$.}
  \label{fig::reaction}
\end{figure}

In Fig.\ \ref{fig::reaction}, we present an example where $R_0(c)$ is asymmetric about $c=0.5$. In the range of $c$ where $R'_0(c)<0$, the pattern is stabilized (destabilized) when $f>0$ ($f<0$). With $R_0(c)$ skewed to the left, the pattern under a positive reaction rate $R$ is more homogeneous than a negative one of the same magnitude. We start from a random initial condition and consider two sets of snapshots where $R=0.08$ and $-1$, respectively.

In Fig.\ \ref{fig::reaction}, we compare the inversion results based on image data from a single or both directions. The reservoir chemical potential $\mu_\text{res}(t)$ is unknown. We find that the systematic error between MAP and the truth is large when only images reacting in a single direction are used as the training data, even though the objective function $S$ is sufficiently small (root-mean-squared error of frames 2--5 is less than 0.5\%). MAP becomes almost identical to the truth when both directions are used, and the strongly concentration-dependent reaction kinetics $R_0(c)$ can be captured. This effect is also reflected in the marginal mean of $\mu_h(c)$ and $R_0(c)$ from the MCMC result. The uncertainty of both functions are significantly reduced when both datasets are used, highlighting the necessity of datasets at different operating conditions in order to infer both thermodynamic and kinetic properties for a chemically driven system whose only observed information is the concentration field (assuming $\mu_\text{res}$ is unknown).

\begin{figure}
  \centering
  \includegraphics[width=0.5\textwidth]{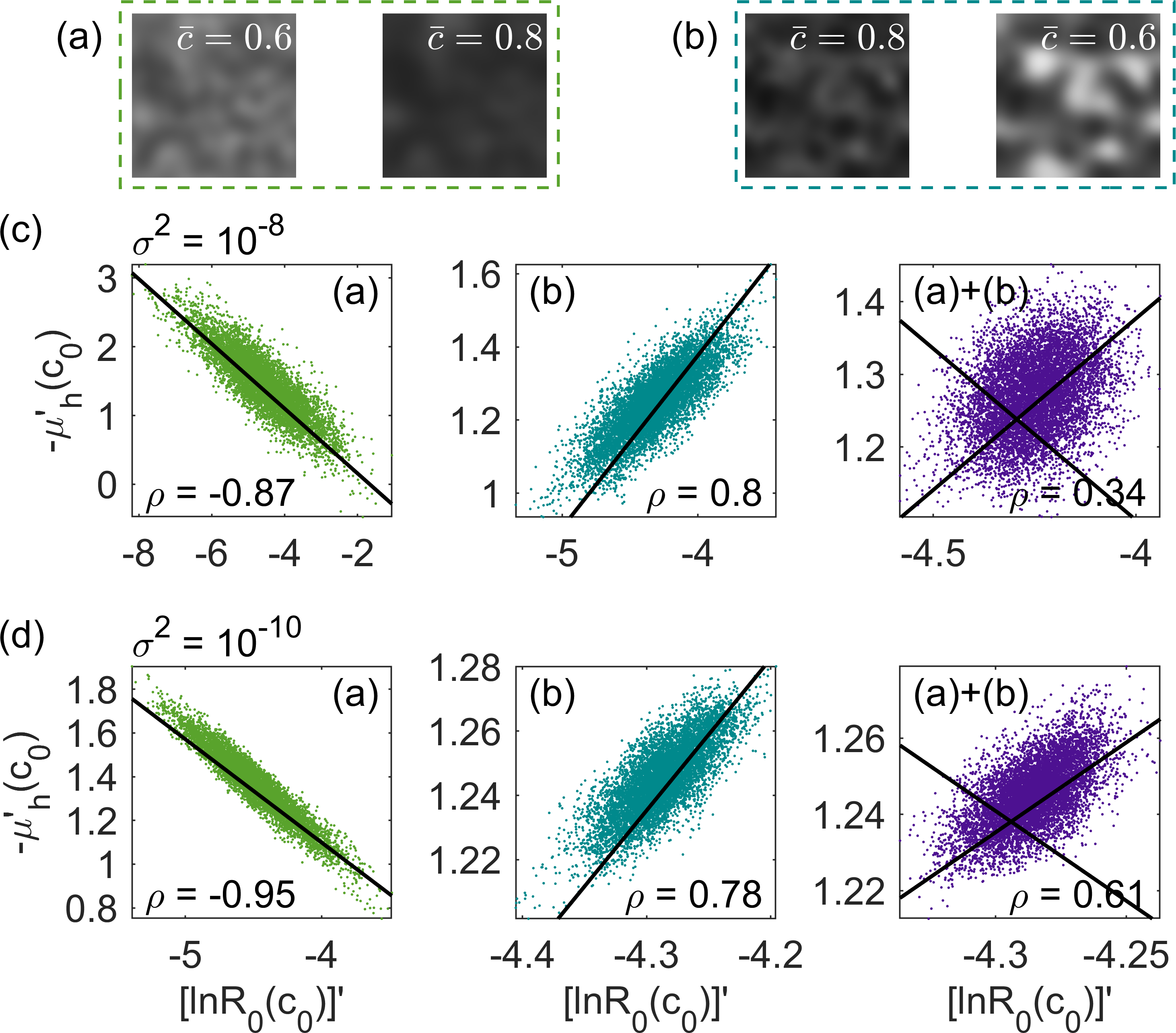}
  \caption{The scatter plots of $\left[\ln{R_0(c_0)}\right]'$ and $-\mu_h'(c_0)$ from the MCMC sample (c and d, where $\sigma^2=10^{-8}$ and $10^{-10}$, respectively) based on datasets (a), (b) and both combined, $c_0=0.7$. The autocatalytic rate $s$ is constant on the solid lines (at $c_0$ and the reaction rate $R$ that corresponds to the dataset used), which analytically predict the correlation observed between $\left[\ln{R_0(c_0)}\right]'$ and $-\mu_h'(c_0)$.}
  \label{fig::reaction_corr}
\end{figure}

Fig.\ \ref{fig::reaction}f shows the scatter plots of $\left[\ln{R_0(c_0)}\right]'$ and $-\mu_h'(c_0)$ ($c_0=0.5$) from the MCMC sample. Using the dataset with $R>0$ ($R<0$), the two quantities are negatively (positively) correlated. When both datasets are used, their correlation is reduced. The correlation can be understood from the dispersion relation mentioned above. Linearizing \ref{eqn::ACR} around a uniform field $c(x)=c_0$, we obtain \cite{Bazant2017,Zhao2019}
\begin{equation}
  \begin{split}
    \omega(k) &= \left. \left[ R_0' f - R_0 f' \cdot \left( \mu_h' + \kappa k^2 \right) \right] \right|_{c=c_0}\\
    &= \left. R \left( \frac{d\ln{R_0}}{dc} - \frac{d\ln{f}}{dc} \cdot \left( \mu_h' + \kappa k^2 \right) \right) \right|_{c=c_0}
  \end{split}
\end{equation}
That is, if $c(x,t=0) = c_0 + \nu e^{i\mathbf{kx}}$, where $\nu \ll 1$ is a small perturbation, then $c(x,t) = c_0 + Rt + \nu e^{\omega(k) t + i\mathbf{kx}}$. Note that $R \to R_0(c)f\left( \mu_\text{res}-\mu_h(c_0) \right)$ as $\nu \to 0$.
Therefore, in the limit of $t\to 0$ and $\nu \to 0$, when multiple or a range of wavenumber $\mathbf{k}$ exists, only $s/R = \left(\ln{R_0}\right)'-\left(\ln{f}\right)' \mu_h'$ and $\left(\ln{f}\right)'$ at $c_0$ can be inferred from the pattern. Note that $s$, also known as the autocatalytic rate \cite{Bazant2017}, is the key to determining the heterogeneity of the pattern: when $s>0$, the pattern becomes linearly unstable and vice versa. Using Butler-Volmer kinetics,
\begin{equation}
  \begin{split}
    \frac{d\ln{f}}{dc} &= \frac{1}{f}\sqrt{\left(\frac{f}{2}\right)^2+1} \\
    &= \frac{1}{R} \sqrt{\left(\frac{R}{2}\right)^2+R_0^2}.
  \end{split}
\end{equation}
Therefore, when $R>0$ ($R<0$), $\left(\ln{R_0}\right)'$ and $-\mu_h'$ are negatively (positively) correlated. To further demonstrate the correlation between the two important quantities, we perform MCMC using two datasets, each of which contains two snapshots with an average concentration of $0.6$ and $0.8$, with $R=0.2$ and $R=-0.2$. In Fig.\  \ref{fig::reaction_corr} with very small observation noise $\sigma^2=10^{-8}$ and $10^{-10}$, we show that $\left[\ln{R_0(c_0)}\right]'$ and $-\mu_h'(c_0)$ ($c_0=0.7$) are strongly correlated with only one dataset used and much weaker correlation is observed when both are used.
The solid lines are $\left(\ln{R_0}\right)'-\left(\ln{f}\right)' \mu_h' = \text{const}$, where the constant and $\left(\ln{f}\right)'$ are determined by the known truth. The agreement of the uncertainty quantification with the analytical analysis demonstrates that when only one dataset obtained under a given reaction rate is available, reaction kinetics and thermodynamics are strongly coupled and only autocatalytic rate $s$ can be determined. To infer both quantities separately (to reduce the correlation of their posterior distribution), datasets under different reaction rates, preferably in different directions, are necessary. Fig.\  \ref{fig::reaction_corr_mu_R0_dmu0} confirms that $R_0$ has a small posterior variance and is not correlated with $\mu_h'$, since $\left(\ln{f}\right)'$ can be determined independently.

\subsection{Temporal resolution}
\label{sec::temporal_res}
\begin{figure}
  \centering
  \includegraphics[width=0.5\textwidth]{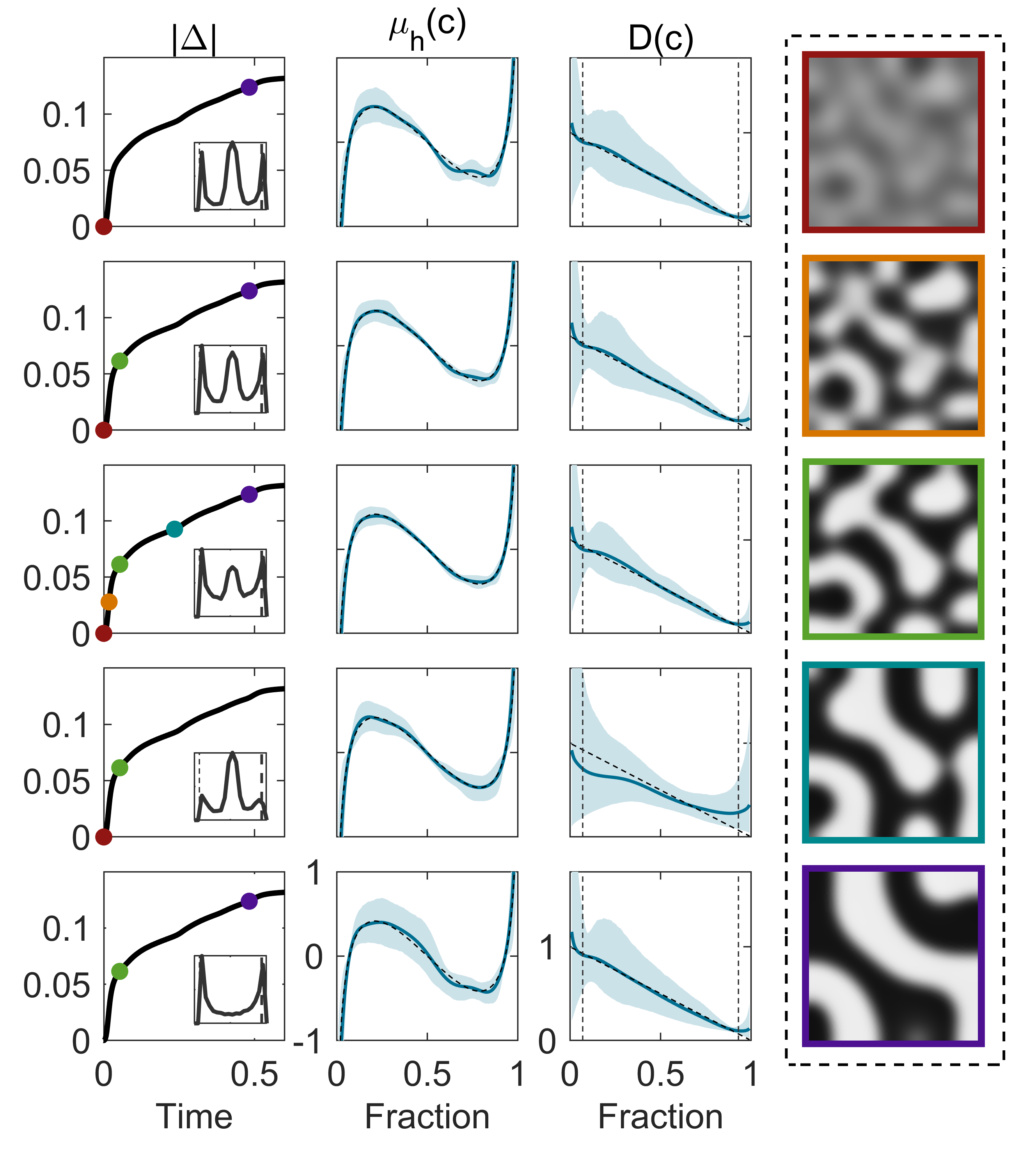}
  \caption{The uncertainty in the chemical potential and diffusivity given snapshots taken at different time over the course of spinodal decomposition. The left columns show the $L_2$ norm of the difference between a pattern at time $t$ and that at $t=0$ ($|\Delta|$). The snapshots included in each row are highlighted as dots in the $|\Delta|-t$ plots whose color corresponds to that of the image outline in the far right column. From the top row to the bottom, 2, 3, and 5 snapshots equally spaced in $|\Delta|$, 2 early-stage snapshots, and 2 late-stage snapshots are taken as the training data, respectively. The insets show the histograms of combined training data of each case.}
  \label{fig::temporal}
\end{figure}

The availability of snapshots over the course of spinodal decomposition and coarsening determines the uncertainty in the inferred parameters. Fig.\  \ref{fig::temporal} shows that with increasing number of snapshots (2, 3, and 5), the uncertainty decreases. These snapshots are evenly spaced in terms of the $L_2$ norm of the difference from the first snapshot. In Fig.\  \ref{fig::temporal} we define the distance between patterns to be $|\Delta|=\int{\left(c(t,\mathbf{r})-c(t=0,\mathbf{r})\right)^2 d\mathbf{r}}$.

At the late stage of coarsening, most pixels in the image are found to be near the miscibility gap $c=c_1$ and $c_2$ (see histograms in Fig.\ \ref{fig::temporal}), and the coarsening rate is most determined by diffusivity near $c_1$ and $c_2$. Whenever a late-stage snapshot is provided, we observe a local minimum in the uncertainty of $D(c)$ at $c_1$ and $c_2$, as highlighted by the vertical dashed lines. If only the early spinodal stage images are provided, the uncertainty for both $\mu_h(c)$ and $D(c)$ away from the initial concentration $c_0=0.5$ increases. Without the early-stage snapshots, the uncertainty of the chemical potential within the miscibility gap is high. While the value of $\mu_h(c)$ at each $c$ may be uncertain, the mean plus and minus 2 standard deviation of $\gamma/\gamma_\text{truth}$ is $1.02 \pm 0.11$, $1.01 \pm 0.065 $, $1.00 \pm 0.056$, $1.02 \pm 0.09$, $1.03\pm 0.10$, respectively, which indicates that two late-stage snapshots are as useful as two early-stage snapshots in providing information about the interfacial tension.

In fact, it is known in phase field theory that the chemical potential of a sphere of radius $R$ is $\gamma/R$ (plus some constant), its growth rate is proportional to its difference with the exterior chemical potential and the detailed functional form of $\mu_h(c)$ is not important. Therefore, images of the dynamics within the miscibility gap are critical in measuring the free energy of a phase-separating system.

\subsection{Spatial resolution} \label{sec::space_res}
\begin{figure}
  \centering
  \includegraphics[width=0.5\textwidth
  \centering]{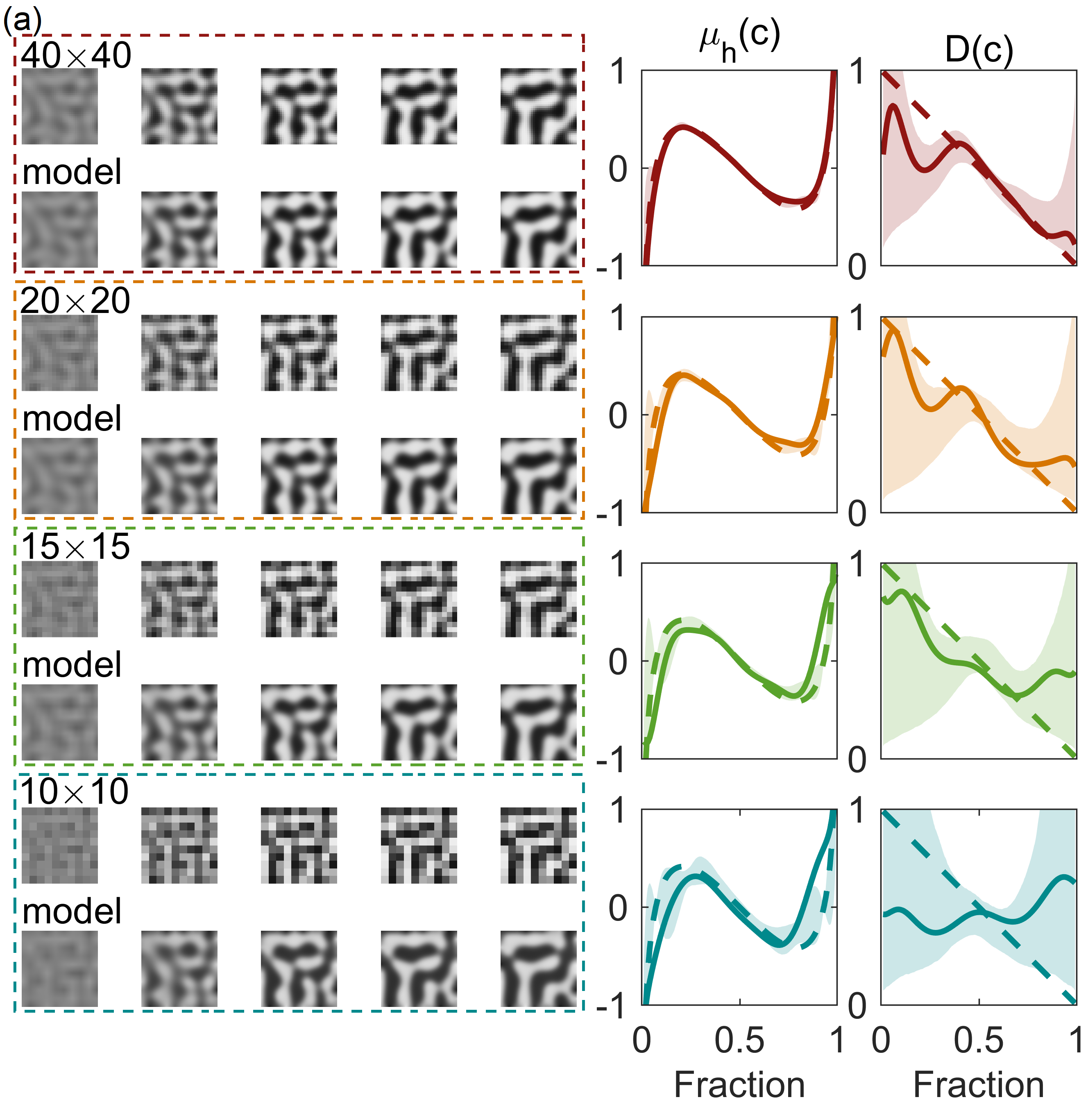}
  \includegraphics[width=0.5\textwidth]{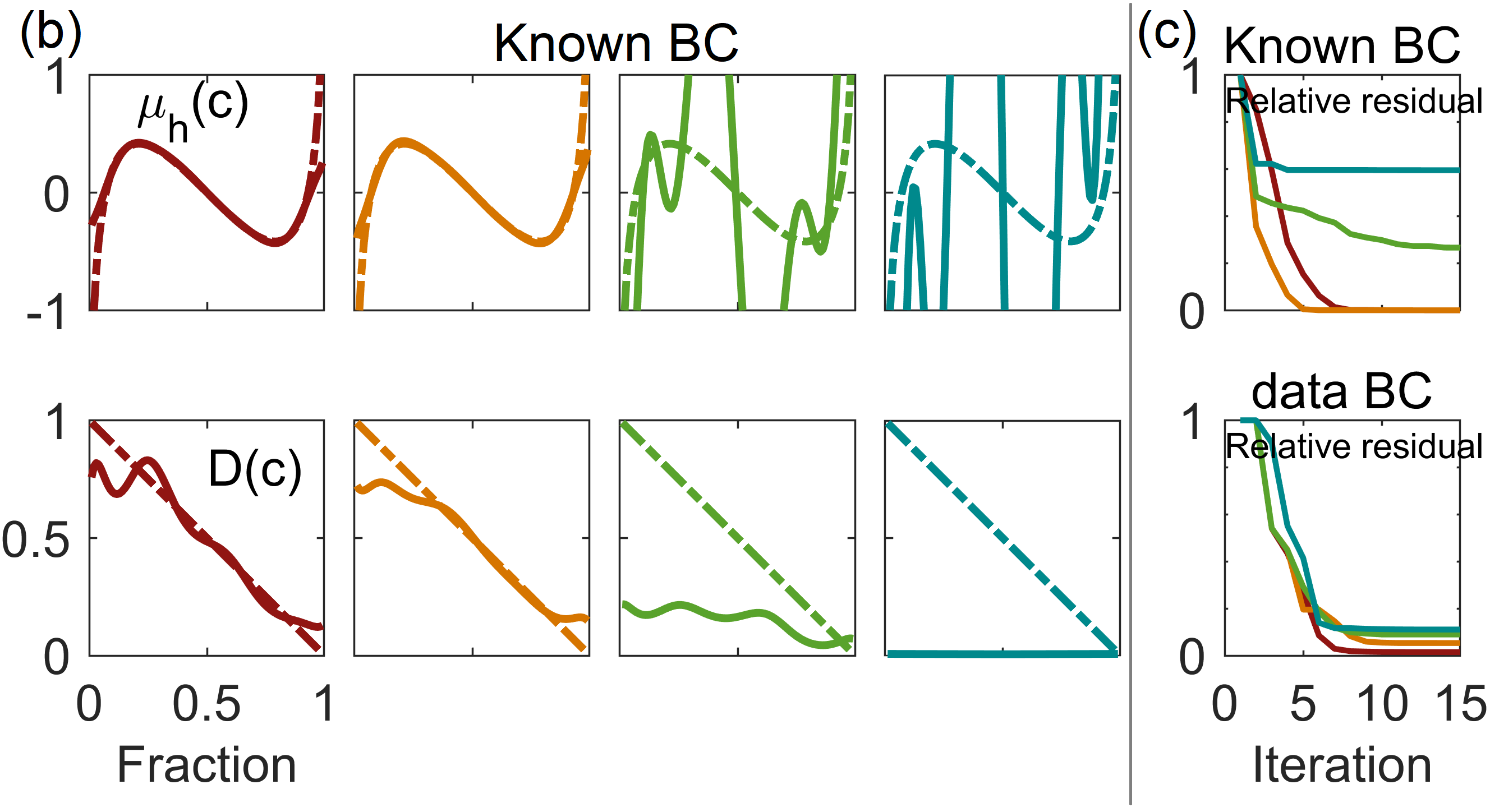}
  \caption{The effect of image resolution on inversion and uncertainty. (a) Using images of different resolutions shown on the right and their interpolated values as the initial and boundary conditions, inversion is performed using boundary condition from the data. The MAP is shown as solid curves in the corresponding right panels. The uncertainty is shown in the same set of plots as the shaded regions. (b) Results of inversion performed on images of the same resolutions (colored consistently) using the known boundary conditions. (c) The residual during training with known boundary conditions (known BC) and boundary condition from the data (data BC).}
  \label{fig::spatial}
\end{figure}

To study the effect of spatial resolution, images are taken from a subset of the field and down-sampled on a rectangular grid. The PDE is numerically solved on a finer grid to resolve the fine details. The first snapshot is linearly interpolated onto the finer simulation grid to be the initial condition. Since the boundary condition for the subset is unknown, all images are also interpolated in space to yield the concentration and its normal gradient at the boundaries, which are then interpolated in time and used as the boundary condition. Assuming that the noise at each pixel and snapshots are independent, the objective function is defined as
\begin{equation}
  S(\mathbf{p}) = \frac{1}{2\sigma_p^2} \left[\sum_{i,j}{ \left( c(t_i,\mathbf{r}_j; \mathbf{p}) - c_{\text{data}}(t_i,\mathbf{r}_j) \right)^2} \right] + \| \mathbf{p} \|_L
\end{equation}
where the pixel-wise variance is $\sigma_p^2 = 10^{-2}$. The likelihood function is defined similarly. Fig.\ \ref{fig::spatial}a shows the MAP as well as the uncertainty from MCMC with different resolutions. Note that we place no constraint on $\mu_h(c)$ for finding the MAP but we fix the miscibility gap for MCMC.

Since the initial condition and boundary conditions are inaccurate at low temporal and spatial resolutions,  MCMC predicts a systematic error compared to the truth. We also compare the MAP result when we enforce the known boundary condition (zero flux and no surface wetting) in Fig.\ \ref{fig::spatial}b. In this case, the optimizer fails to get even close to the truth at low resolution. Therefore, when the spatial resolution is low, it is preferable to initialize from the first snapshot and impose boundary conditions from the data itself, despite the fact they are not accurately known. The inaccuracy in the initial condition can be partially compensated by the boundary condition. In fact, comparing the model prediction from the MAP results at different resolutions, the patterns are largely preserved. The optimizer fails to find a reasonable solution at a resolution of $8\times 8$. Therefore, as a rule of thumb, at least 3 pixels per wavelength of spinodal pattern are needed.

Without constraining the miscibility gap, the optimizer converges to a $\mu_h(c)$ with smaller miscibility gap with decreasing spatial resolution. This occurs because the low sampling rate effectively filters out the high-frequency components, blurring the high contrast between the two phases. 

\subsection{Image domain size}
\label{sec::domain_size}
\begin{figure}
  \centering
  \includegraphics[width=0.5\textwidth]{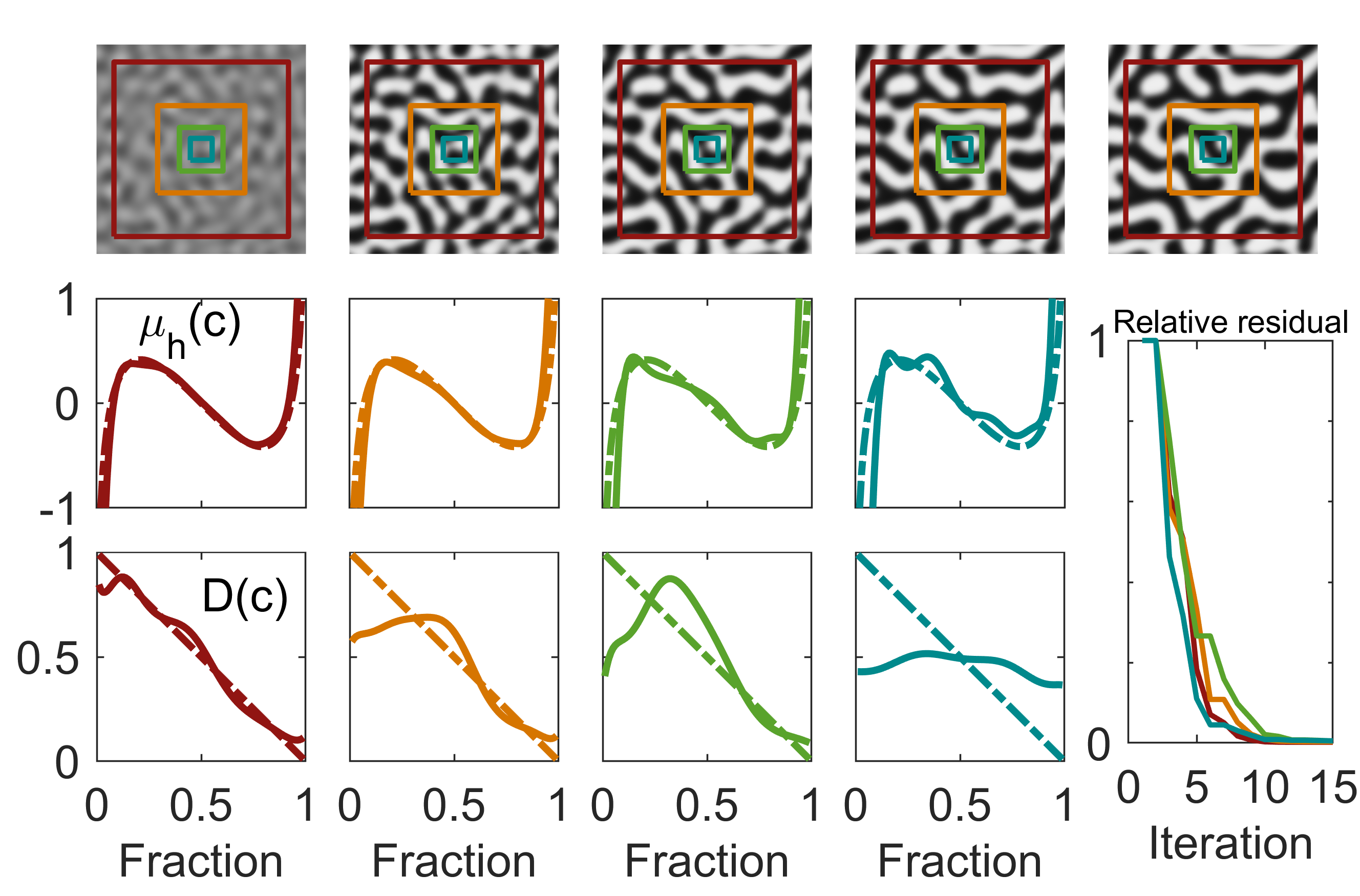}
  \caption{Inversion performed on images of different physical sizes. Initial and boundary conditions are imposed from data interpolated in time. The chemical potential and diffusivity are the MAP results based on training data outlined by the same color. The residual plot of all cases are shown on the right.}
  \label{fig::size}
\end{figure}

In cases where the field of view is limited to a subset of the entire domain, the boundary condition is unknown. Similar to Section \ref{sec::space_res}, the concentration and its normal gradient from each snapshots are interpolated in time and used as the boundary conditions. Fig.\ \ref{fig::size} shows the training result for images of different domain sizes. The discrepancy between the truth and the MAP increases with decreasing domain size, suggesting that with the parameters become less identifiable with less information and local minima of the objective function is likely to be encountered. Only lower order polynomials of $D(c)$ can be inferred when the domain size are too small to contain large concentration variations. However, the computational cost for solving the PDE is greatly reduced if only a smaller subdomain in the training data is used. The regularization parameter is $10^{-5}$ times the domain size.

\subsection{Blurring filter}
\label{sec::blurring}
Imaging systems have a certain point source function (PSF) that may spread over more than a single pixel. The images are a convolution of the object and the PSF. Therefore studying the inversion of these blurred images is important for practical imaging devices. In the example of spinodal decomposition used above, we show that using blurred images directly will result in systematic error in the inferred chemical potential and diffusivity. However, it is possible to invert the characteristic length scale of the PSF and more accurate physical properties simultaneously, effectively leading to a physics-constrained deconvolution.

\begin{figure}
  \centering
  \includegraphics[width=0.5\textwidth]{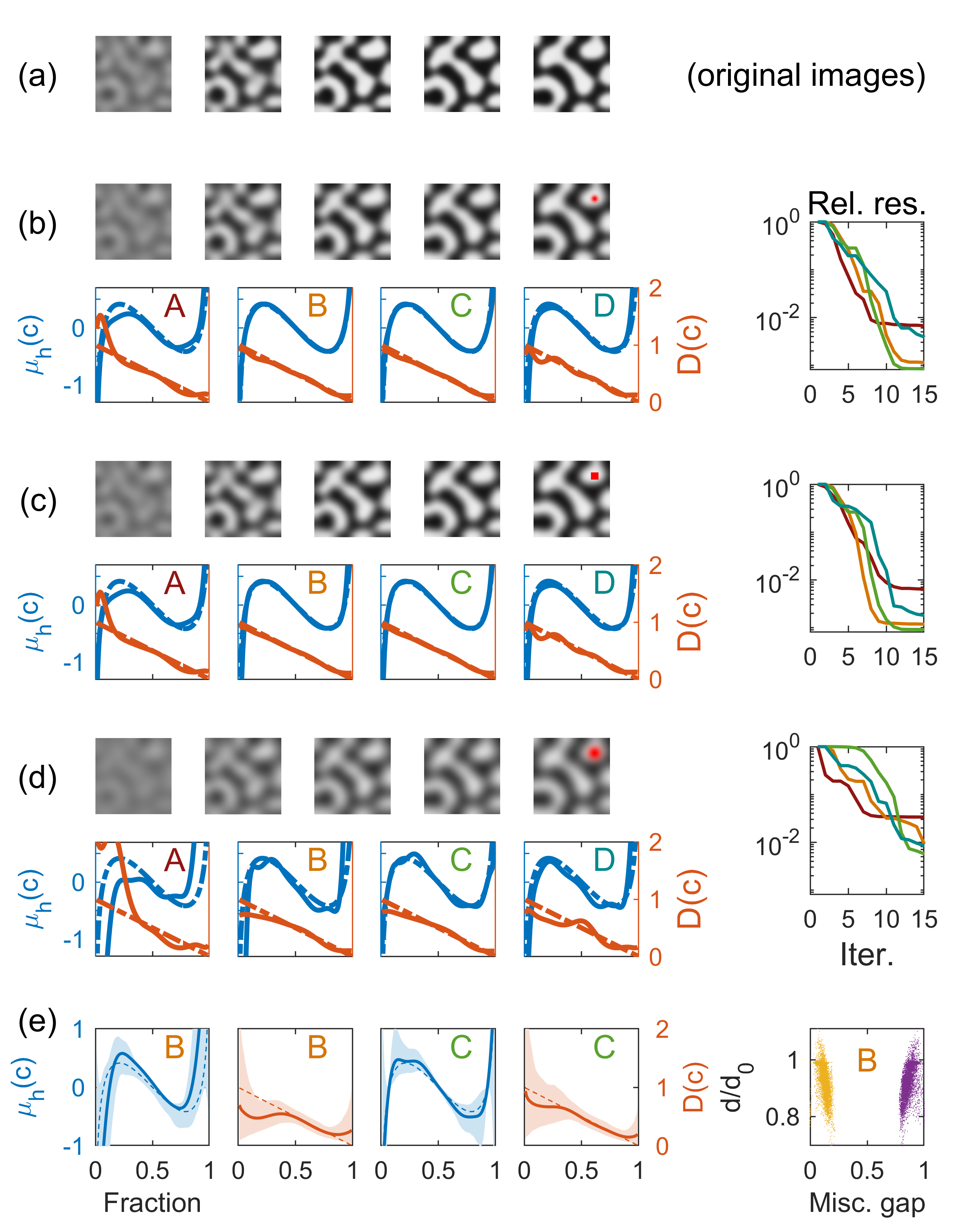}
  \caption{Inversion performed on blurred images. (a) original images; (b,c,d) Inversion of $\mu_h(c)$ and $D(c)$ based on blurred images whos PSFs are shown in red; (e) Uncertainty quantification based on images in (c). Letters A--D correspond to inversion strategies described in the main text. $\mu_h(c)$ are plotted in blue and $D(c)$ are plotted in orange. Solid lines are the MAP solutions in (b)--(d) and marginal mean in (e). Dashed lines are the known truth. ``Rel.\ res.'' stands for relative residual. The scatter plot shows the MCMC sample of the Gaussian filter length $d$ and the miscibility gap $c_1$ and $c_2$.}
  \label{fig::blur}
\end{figure}

In Fig.\ \ref{fig::blur}, we generate a sequence of images in (a) and convolve them with Gaussian (b and d) and box-averaging PSF (c). The PSF is indicated by the red region in the upper right corner of the last image in the sequence. The Gaussian PSF is common and can be used to approximate an Airy disk, which is the PSF of a circular aperture. Box-averaging PSF is constant within a compact support. We adopt four inversion strategies, denoted in Fig.\ \ref{fig::blur} as A) assume the images are not blurred; B) assume the images are blurred by a Gaussian PSF and invert its standard deviation $d$ together with $\mu_h(c)$ and $D(c)$; C) the same as B, except that $\mu_h(c)$ is constrained to a fixed miscibility gap, which may be measured more accurately after a long relaxation into two well-separated phases; D) the same as B, except that the first image is deconvolved and then used as the initial condition. In strategies A--C, the first given image (blurred) is used as the initial condition. In all strategies, the regularization parameter is $10^{-2}$, the objective function is a discrete summation of the squared error on five $50\times 50$ images. The initial guess is $d/L = 0.1$, $\alpha = 0.1$ (noise-to-signal ratio for deconvolution, see below), $D(c) = 0.1$.

Strategy A (treating blurred images as the truth) underestimates the interfacial tension and the miscibility gap in all cases. The objective function value that the optimizer converged to is also significantly higher than other strategies. This effect becomes more severe when the length scale of the PSF increases and becomes comparable to the spinodal length scale, $l = \sqrt{2\kappa/\mu_h'(c_0)}$. The standard deviation of the Gaussian PSF is $d=0.03L=0.67l$ and $d=0.06L=1.34l$ respectively for Figs.\
\ref{fig::blur}b and d, where $L$ is the image size. For the latter case, $[-3d,3d]$ extends to about one wavelength of the spinodal pattern $2\pi l$. The side length of the averaging box in Fig.\ \ref{fig::blur}c is $0.1L$.

When the PDE solution is convolved with a Gaussian PSF and $d$ is also optimized (strategy B), the inferred $\mu_h(c)$ and $D(c)$ are closer to the truth. For the sets of images blurred by a Gaussian PSF, the correct $d$ is found, while for Fig.\ \ref{fig::blur}c the solver converges to $d/L=0.03$ .
For images whose unknown PSF is bounded and decays to zero, the Gaussian PSF is often a good estimate. Here, nonzero Fourier components of the images are concentrated in a narrow band around $l^{-1}$; blurring is only sensitive to the characteristic length scale of the PSF and not its details. Fig.\ \ref{fig::blur}c shows that an approximation with a Gaussian PSF is sufficient. In addition, Gaussian PSF is differentiable with respect to its parameter $d$, hence useful for gradient-based optimizers. Given the physical constraint of miscibility gap, strategy C shows that the objective function is further decreased.

Strategy D deconvolves the first image to use as the initial condition. To reduce the ringing effect of deconvolution due to discontinuity at the boundary, we perform a linear interpolation between the given image and the blurred image convolved with the PSF, weighing the given image more in the interior and the blurred image more at the boundary (to reduce the high frequency components near the boundary). We use $\left(1-e^{-x^2/4d^2}\right) \left(1-e^{-y^2/4d^2}\right)$, shifted so that $[x,y]=[0,0]$ corresponds to the corner. We use Wiener deconvolution which in the Fourier space is $K^* \left( K^*K+\alpha \right)^{-1}$, where $K$ is the PSF in Fourier space, $\alpha$ is the noise-to-signal ratio and included as a variable to be optimized. The residual plots and plots of $\mu_h(c)$ and $D(c)$ show that this strategy is close to or sometimes worse than strategies $B$ and $C$ which do not deconvolve the first image, which shows that, while the initial condition can be polluted by noise and blurring, preprocessing may be unnecessary. In fact, the unknown boundary condition may introduce additional error in the process of deconvolution. Inversion of a PDE whose solution diverges when the initial condition is slightly perturbed \cite{Pathak2018} is beyond the scope of this work.

The uncertainty using strategies B and C based on images in Fig.\ \ref{fig::blur}c and $\sigma_p^2 = 0.1$ is summarized in Fig.\ \ref{fig::blur}e. The scatter plot shows a strong and negative correlation between the inferred Gaussian filter length $d$ and the width of the miscibility gap when $\mu_h(c)$ does not have any constraint, further demonstrating the importance of imposing the physical constraint. When such a constraint is imposed, the standard deviation of $d/d_0$ is reduced from 0.07 to 0.035, where $d_0$ is the truth. The miscibility gap $[c_1,c_2]$ is defined by Eq.\  \ref{eqn::mu_constraint} together with $\mu'_h(c_1)>0$, $\mu'_h(c_2)>0$ and $c_1 \neq c_2$.

\section{Conclusion and outlook}
In summary, using the approach of PDE-constrained optimization and Bayesian inference, we performed a systematic analysis of the inversion and the uncertainty quantification of the constitutive relations based on images of pattern formation.
We showed that the optimization is robust with the use of estimated Hessian, and that the MCMC sampling is efficient in a high dimensional parameter space. For a phase-separating system, the posterior distributions of free energy (or the corresponding interfacial energy) and diffusivity based on the image data are uncorrelated, which is also corroborated by scaling analysis.
The uncertainty in the inferred quantities depend strongly on the average composition and the relative contribution of reaction and diffusion.
For a chemically reactive system that is driven out of equilibrium, a linear stability analysis confirms the MCMC result that the reaction kinetics and free energy can be inferred separately only when images generated at multiple reaction rates or directions are available.
We show that images taken at earlier stages of spinodal decomposition can be used to infer the functional form of nonlinear free energy and diffusivity more accurately than those from the coarsening process, while the latter can also inform the interfacial tension sufficiently accurately. When the spatial resolution of the images is low, the image domain is smaller than the physical domain, or when the boundary condition is unknown, the inversion remains robust by imposing the data at the image boundary as the boundary condition in the simulation, while the uncertainty increases with decreasing spatial resolution and domain size.
When the images are blurred, the inversion algorithm can extract the constitutive relation as well as the characteristic size of the blurring kernel using the first frame as the initial condition without deblurring, yet there exists a correlation between the inferred miscibility gap and the kernel size. Including the prior knowledge of the miscibility gap as a constraint further reduces the uncertainty of the inferred quantities.
The examples show that when the imaging quality is limited by the instrument, such a PDE-constrained inversion effectively becomes a physics-informed superresolution imaging technique.

The methods and applications discussed here serve as a first step toward quantitative frame-by-frame and pixel-by-pixel matching between experiments and theoretical models, as excellent agreement has already been observed in many complex systems \cite{Berry2015,Wensink2012,Dunkel2013,Liu2019}. Despite the lack of microscopic information, the macroscopic pattern-formation dynamics can be described parsimoniously by PDEs with relatively few parameters. The inversion of the PDE means that, from the images, these macroscopic physical properties can be measured that would otherwise be inaccessible especially for systems far from equilibrium such as active matter and biology, where free energy is ill-defined, and nonequilibrium thermodynamics is poorly understood. The uncertainty quantification can be applied to optimal design of experiments to carefully probe regions of higher uncertainty as informed by prior experimental data.

The inversion also enables a physically interpretable parametrization of complex systems, which may help in establishing a mapping between the macroscopic and microscopic parameters, and eventually engineering or controlling patterns by tuning physical properties of the constituents, as reported recently in the biological engineering of Turing patterns \cite{Kryuchkov2020} and the design of PDEs to create desired patterns \cite{Yoshinaga2020}. Our computational approach can be integrated into the loop to accelerate the search in a high dimensional parameter space by identifying the most important engineering handles.

While phase field (Cahn-Hilliard and Allen-Cahn models) and phase field crystal models were selected as the model systems in this article, the approach can be readily extended to other systems, such as for fluid dynamics and multi-component reaction-diffusion equations, where more complicated patterns may arise \cite{Cross1993,Chandrasekhar1981,Pearson1993}, and the sensitivity of patterns and bifurcation dynamics with respect to constitutive relations in a high-dimensional parameter space awaits exploration. For complex systems, further study is needed to quantify the range of phenomenon that a model can describe. Techniques in inverse problems, dynamical systems, and identifiability analysis should be employed when discrepancy between experiments and models arise due to nonidealities such as spatial heterogeneity as well as unknown hidden variables.

\renewcommand{\thefootnote}{\alph{footnote}} 
\renewcommand{\theequation}{S\arabic{equation}} 
\renewcommand{\thefigure}{S\arabic{figure}} 
\renewcommand{\thetable}{S\arabic{table}} 
\renewcommand{\thesection}{S\arabic{section}} 

\appendix
\section{Comparison of optimization and uncertainty quantification algorithms}
\label{sec::appendix_alg}
Similar to the generation of Fig.\ \ref{fig::ensemble}, we generate 100 realizations of spinodal decomposition snapshots and perform the optimization using either all five snapshots or the first and last snapshots, starting from the same initial guess as described in the main text. The termination criteria is $10^{-6}$ for optimality, function, and step tolerance. The objective function values that the optimizer converges to for different realizations, if at the global minimum, are very close and much smaller than values at local minima, hence can be easily classified. Table.\ \ref{table::alg_cp} lists the percentage of successful convergence to the global minimum, the number of iterations, and computational time for successful cases using FSA (trust-region algorithm) and ASA (gradient descent with BFGS). We can see the ASA performs poorly on every metric considered, although it improves when initial guess is brought closer to the truth. Note that, using the ASA method, each iteration involves a line search, which usually involves multiple function and gradient evaluation, as shown by its longer average computational time per iteration than that of FSA. In FSA, each iteration requires one function, gradient and Hessian evaluation.
\begin{table*}
  \centering
  \caption{A comparison of the performance of FSA and ASA. \#  of snapshots: number of snapshots used as the training data. Success ratio: for 100 different datasets, the percentage of optimizations that converge to the global minimum. Iterations to success: the number of iterations (10\% - median - 90\% percentile) taken by the optimizer to converge to the global minimum (cases that converge to a local minima are not considered). Relative time to success: the median computational time relative to the first row for cases that converge to the global minimum.}
  \label{table::alg_cp}
  \begin{tabular}{c|c|c|c}
    Method and \# of snapshots & Success ratio & Iterations to success & Relative time to success \\ \hline
    FSA, 5 & 75\% & 21-31-43 & 1 \\ \hline
    FSA, 2 & 81\% & 19-25-34 & 0.7 \\ \hline
    ASA, 5 & 5\% & 33-65-93 & 4.6 \\ \hline
    ASA, 2 & 47\% & 66-80-100 & 13.0
  \end{tabular}
\end{table*}

\begin{figure}
  \centering
  \includegraphics[width=0.5\textwidth]{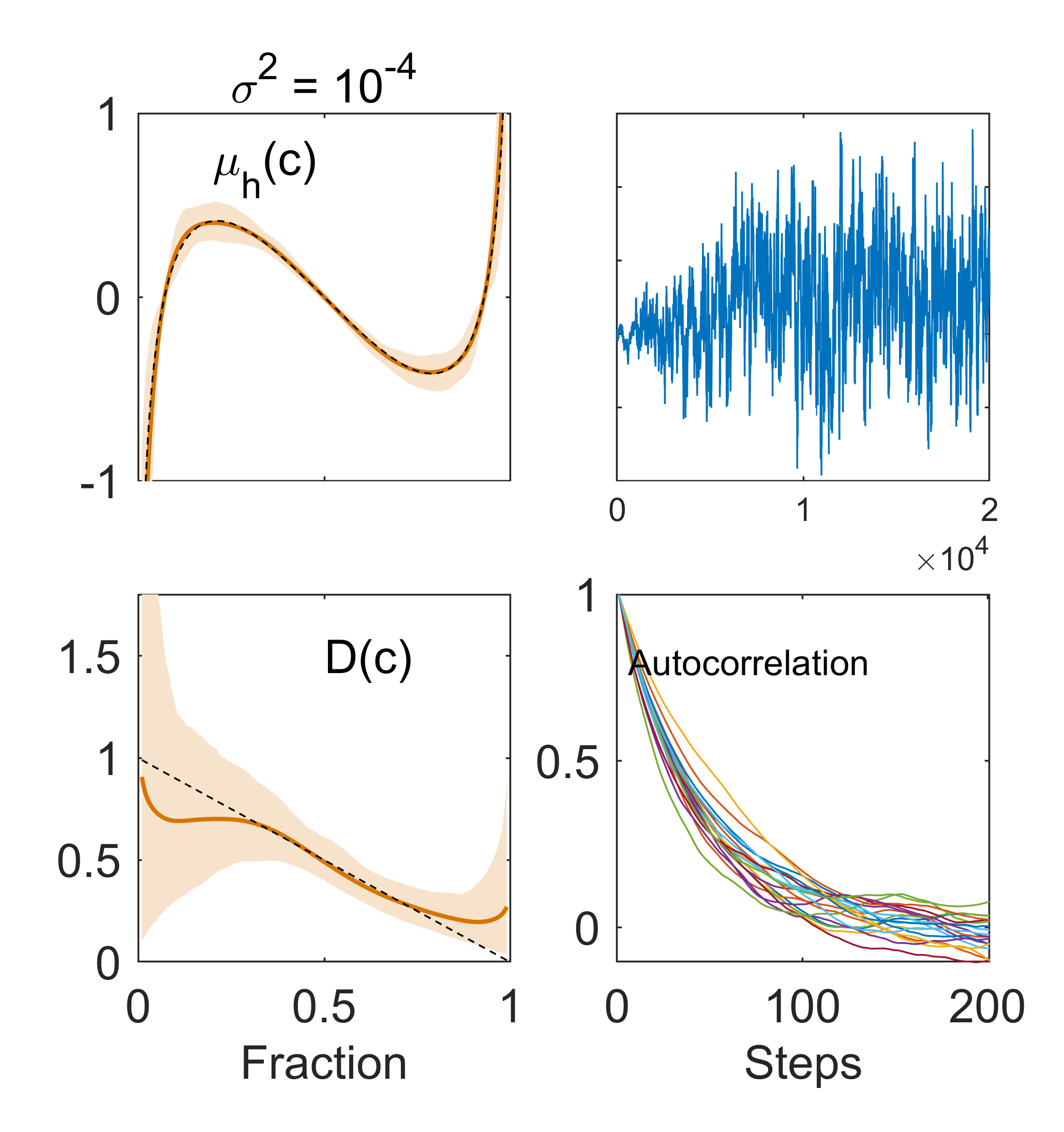}
  \caption{ Uncertainty quantification of the chemical potential and diffusivity from 5 snapshots of spinodal decomposition as shown in Fig.\  \ref{fig::ensemble}. The measurement noise is $\sigma^2 = 10^{-4}$. The shaded regions are the 95\% confidence interval of the functions at each $c$. The solid lines are the marginal mean of the functions at each $c$. The dashed lines are truth. The two panels on the right are the Markov chain trajectory of $a_1$ and the autocorrelation of all 20 parameters, respectively.}
  \label{fig::AM}
\end{figure}

Fig.\ \ref{fig::AM} shows the results of an adaptive MCMC applied to the same problem in Fig.\ \ref{fig::MCMC}. The covariance of the proposal distribution at step $n$ is $C_n = s_d \text{Cov}(X_0,\ldots X_{n-1})+s_d \epsilon I$. We choose $\epsilon=0.001$. Both Fig.\ \ref{fig::AM} and Fig.\ \ref{fig::MCMC} generate 20,000 samples. The uncertainty computed in Fig.\ \ref{fig::AM} discards the first 5000 samples as the burn-in and is sufficiently close to Fig.\ \ref{fig::MCMC}. The autocorrelation is stronger.

\section{Correlation between thermodynamic and kinetic properties}
\begin{figure}
  \centering
  \includegraphics[width=0.5\textwidth]{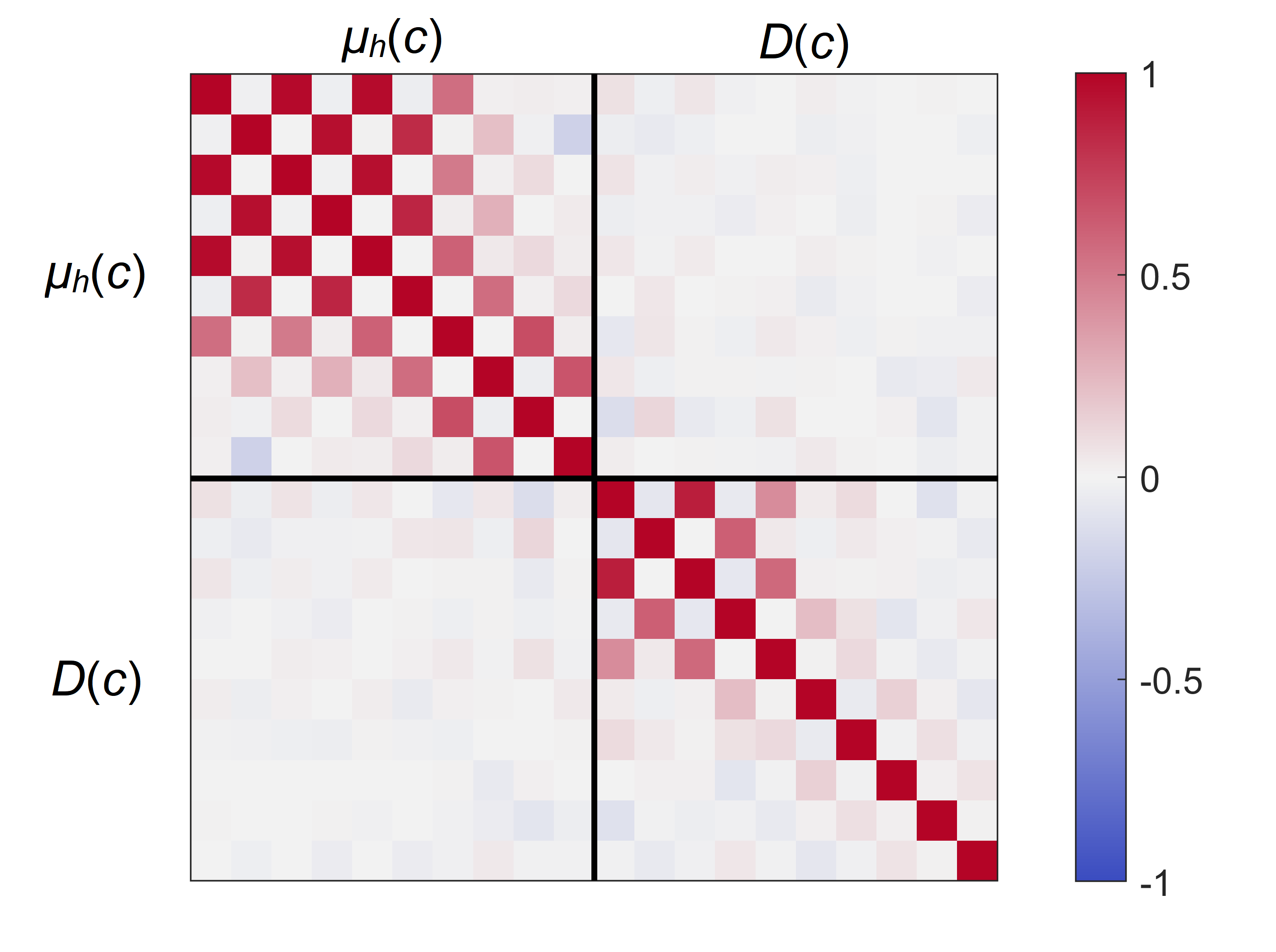}
  \caption{Correlation coefficients between parameters for $\mu_h(c)$ and $D(c)$.}
  \label{fig::corrcoef}
\end{figure}

Fig.\ \ref{fig::corrcoef} shows the matrix of correlation coefficients between parameters for $\mu_h(c)$ ($a_1,a_2,\ldots,a_{10}$)
and $D(c)$ ($b_0,b_1,\ldots,b_9$) inferred from spinodal decomposition snapshots (same as used by Fig.\ \ref{fig::MCMC}) and $\sigma^2=10^{-4}$. The correlation coefficient between $a_i$ and $a_j$ is $\text{Cov}(a_i,a_j)/\sqrt{\text{Var}(a_i)\text{Var}(a_j)}$, where Cov and Var stand for the covariance and variance, respectively.

\begin{figure}
  \centering
  \includegraphics[width=0.5\textwidth]{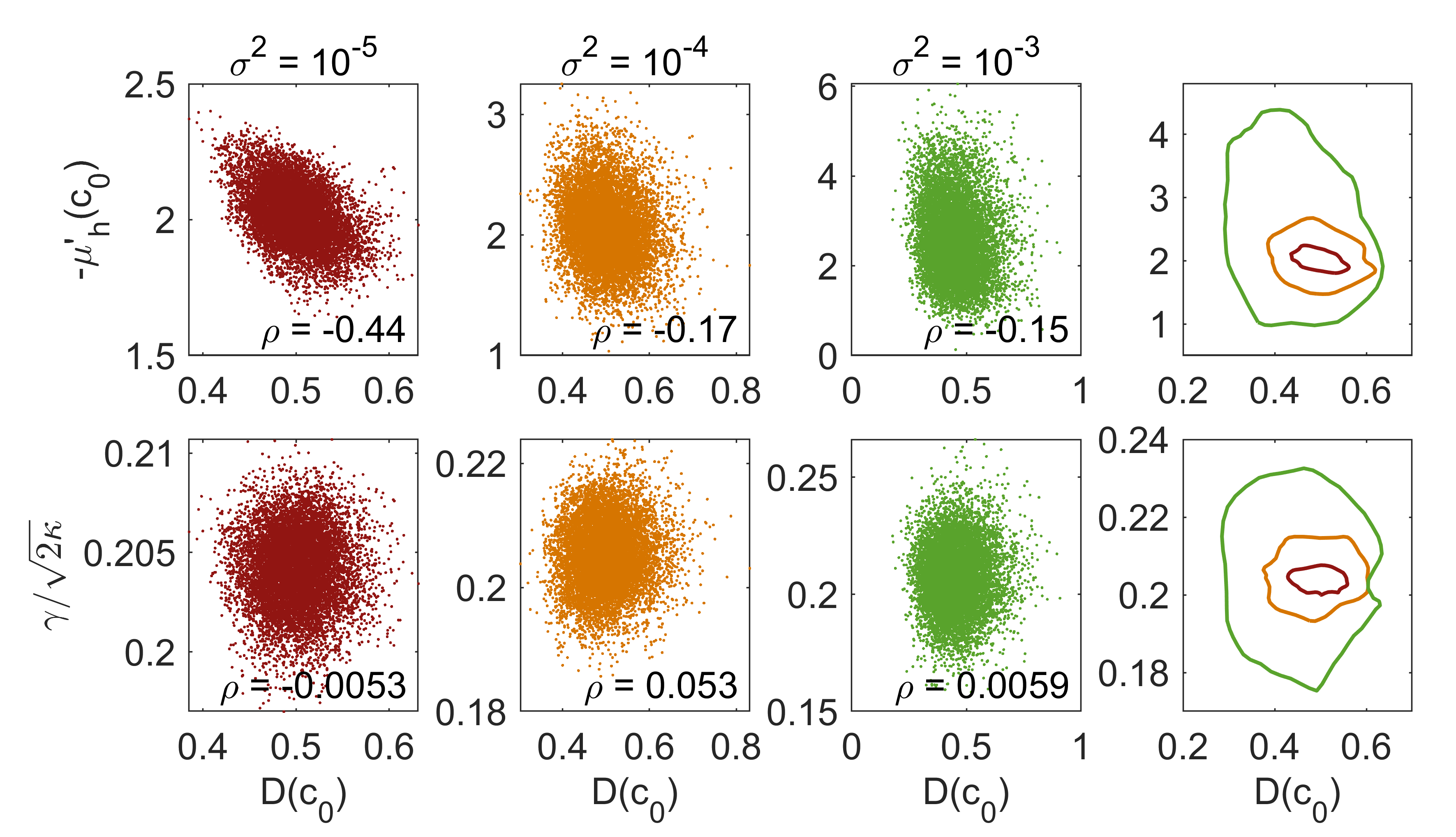}
  \caption{Scatter plots and 90\% confidence region of $-\mu'_h(c_0)$ versus $D(c_0)$ and $\gamma/\sqrt{2\kappa}$ versus $D(c_0)$ (20,000 samples). The colors of the contours match those of the scatter plots.}
  \label{fig::mu_D_corr}
\end{figure}

Using the same set of images, Fig.\  \ref{fig::mu_D_corr} shows the scatter plot of $-\mu'_h(c_0)$ versus $D(c_0)$ from the MCMC chain, as well as $\gamma/\sqrt{2\kappa}$ versus $D(c_0)$, where $c_0 =0.5$ is the average fraction. $\rho$ is the correlation coefficient between the two parameters as defined above. The initial rate of decomposition is $c_0D(c_0)\left[\mu'_h(c_0)\right]^2/4\kappa$, which explains the negative correlation between $-\mu'_h(c_0)$ and $D(c_0)$. The contours are the 90\% confidence region.

\begin{figure}
  \centering
  \includegraphics[width=0.5\textwidth]{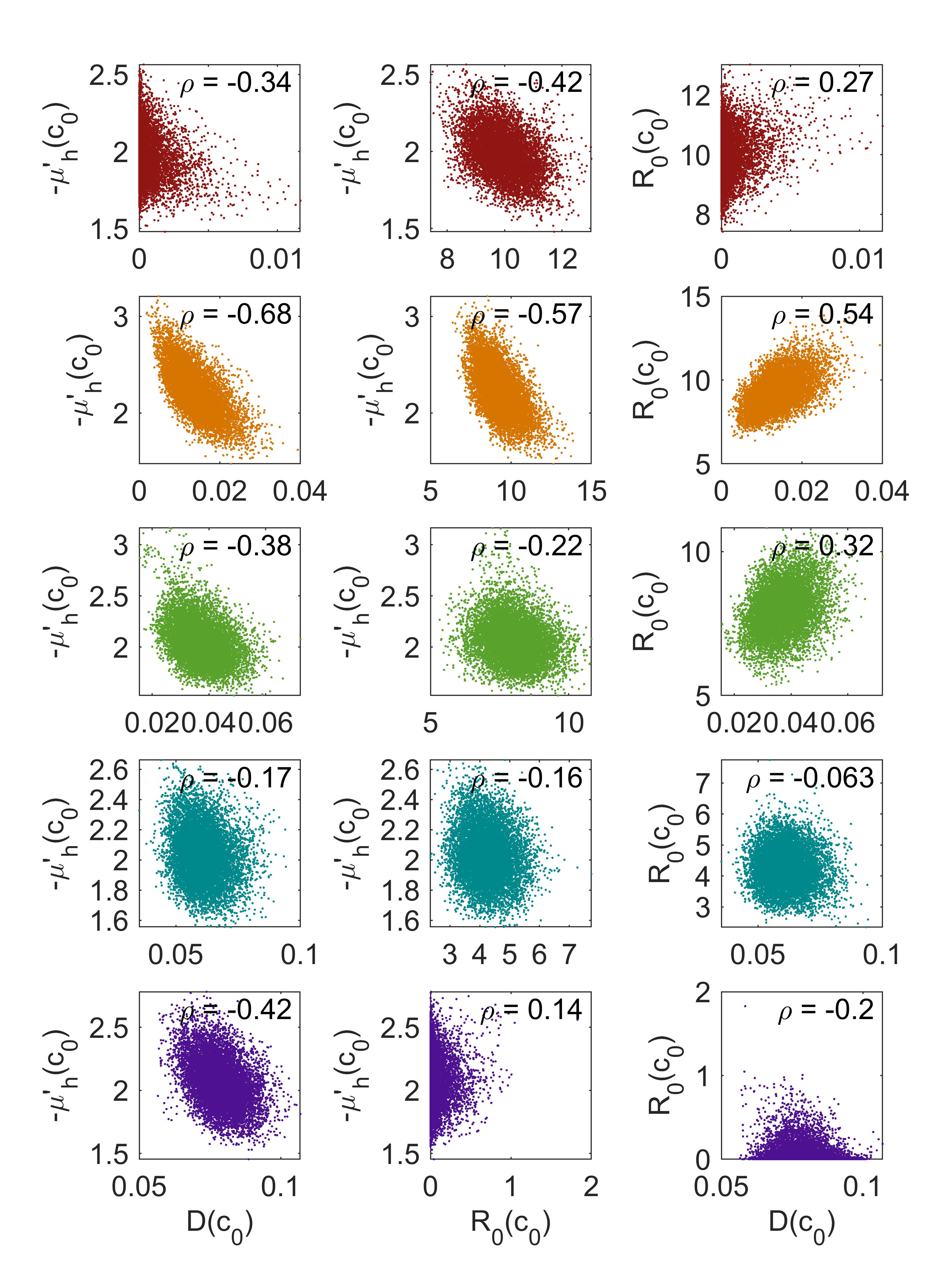}
  \caption{Scatter plots of $-\mu'_h(c_0)$, $D(c_0)$,  and $R_0(c_0)$ (20,000 samples). Each row has a color that corresponds to the set of images having the same color in Fig.\ \ref{fig::model_selection}a.}
  \label{fig::model_selection_corr}
\end{figure}

In a system where both reaction and diffusion are present, we study the pairwise correlation among thermodynamics, reaction kinetics, and diffusivity by showing the scatter plots of $-\mu'_h(c_0)$, $D(c_0)$,  and $R_0(c_0)$ in Fig.\ \ref{fig::model_selection_corr}. The samples are drawn based on images in Fig.\  \ref{fig::model_selection}.

\begin{figure}
  \centering
  \includegraphics[width=0.5\textwidth]{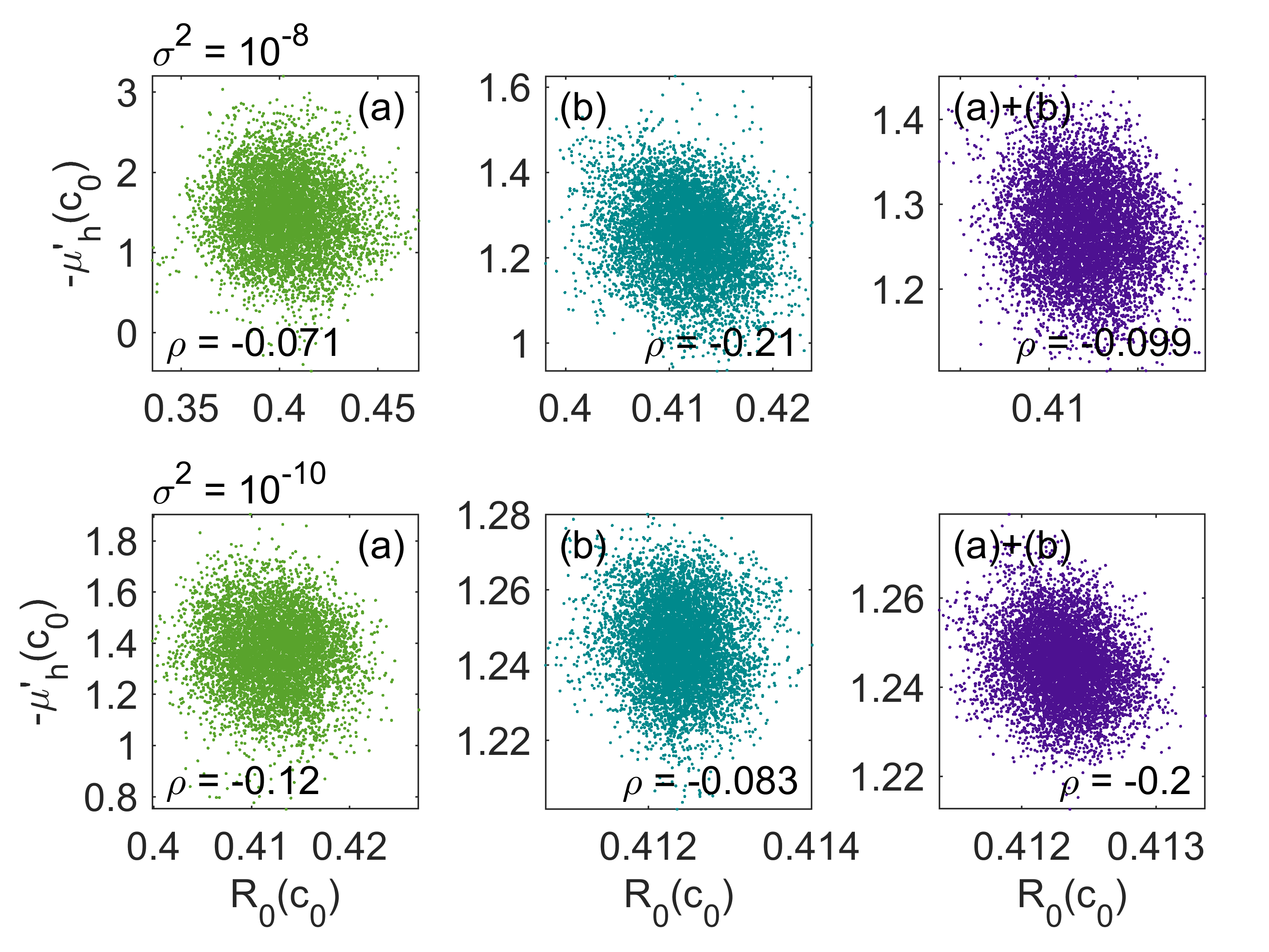}
  \caption{Scatter plots of $R_0(c_0)$ versus $\mu_h'(c_0)$  using datasets in Fig.\  \ref{fig::reaction_corr} (20,000 samples), $c_0=0.7$.}
  \label{fig::reaction_corr_mu_R0_dmu0}
\end{figure}

Using the chemically driven concentration fields studied in Fig.\ \ref{fig::reaction_corr}a and b, Fig.\  \ref{fig::reaction_corr_mu_R0_dmu0} shows the scatter plots of $R_0(c_0)$ and $\mu_h'(c_0)$ where $c_0=0.7$ and confirms that they are uncorrelated and $R_0(c_0)$ can be determined with high accuracy.

\section*{Acknowledgments}
This work was supported by Toyota Research Institute through the D3BATT Center on Data-Driven-Design of Rechargeable Batteries.

\bibliography{Zhao}

\end{document}